\documentclass[superscriptaddress, amsmath,amssymb, aps, twocolumn, nofootinbib]{revtex4-1}

\usepackage{graphicx}
\usepackage{dcolumn}
\usepackage{bm}
\usepackage{wasysym}
\usepackage{hyperref}
\usepackage[mathlines]{lineno}
\usepackage{xcolor}
\usepackage{dsfont}
\usepackage{placeins}
\usepackage[normalem]{ulem}
\usepackage{esint}
\usepackage{hhline}

\usepackage{gensymb}
\usepackage{enumitem}
\usepackage{multirow}
\usepackage{booktabs}
\usepackage{array}
\newcolumntype{P}[1]{>{\centering\arraybackslash}p{#1}}
\usepackage{amssymb}
\usepackage{adjustbox}
\usepackage{orcidlink}
\usepackage{graphicx}  
\newcommand{\iu}{\mathrm{i}\mkern1mu}
\newcommand{\du}{\mathrm{d}}

\DeclareUnicodeCharacter{2009}{-}

\begin{document}

\title{Ray tracing ultracompact boson stars: visibility modulations and incomplete photon rings}

\author{Seppe J. Staelens \orcidlink{0000-0002-1262-1600}}
\email{ss3033@cam.ac.uk}
\affiliation{DAMTP, Centre for Mathematical Sciences,
University of Cambridge, Wilberforce Road, Cambridge CB3 0WA, UK}
\affiliation{Leuven Gravity Institute, KU Leuven,
Celestijnenlaan 200D box 2415, 3001 Leuven, Belgium}

\author{Ece Alkan \orcidlink{0009-0002-3999-9838}}
\email{ea631@cam.ac.uk}
\affiliation{Institute of Astronomy, University of Cambridge, Madingley Road, Cambridge, CB3 0HA, UK}

\date{\today}

\begin{abstract}
Modified black holes and black hole mimickers can give rise to additional photon rings in ray-traced images. With the photon ring structure of Kerr well understood, the question arises how well future versions of the Event Horizon Telescope images would be able to constrain the presence of non-Kerr photon rings. In this work, we investigate how likely additional photon rings are to present themselves through low-frequency modulations of the visibility amplitude, i.e.~in Fourier space where the data live. To this end, we generate ray-traced images of stable and unstable ultracompact, solitonic boson stars, both in spherical symmetry and with rotation. We find that clear modulations in the associated visibility amplitude are only present in spherical symmetry, for a restricted range of observer inclinations, hence suggesting that these modulations can not be a generic expectation of ultracompact black-hole mimickers. Additionally, we discuss how the photon rings are \textit{incomplete} for the rotating boson stars considered, and how an unstable prograde light ring may be needed to remedy this.
\end{abstract}

\maketitle

\section{Introduction}
\label{sec: introduction}

Recent years have seen a strong interest in the study of \textit{exotic compact objects} (ECOs). On the one hand, ECOs are postulated as alternatives to the black-hole (BH) paradigm, circumventing potential theoretical issues like singularities and the information paradox --- see e.g.~Ref.~\cite{Bambi:2025wjx} for a review. On the other hand, ECOs could simply constitute part of the dark matter content of the Universe and coexist with a BH population.\\

While a zoo of ECOs exists (see e.g.~Refs.~\cite{Cardoso:2019rvt, Bezares:2024btu} for reviews), a particularly interesting subset are boson stars (BSs): self-gravitating scalar field configurations that can attain stellar or supermassive mass scales when the scalar is sufficiently light \cite{Kaup:1968zz, Ruffini:1969qy, Friedberg:1986tq, Kleihaus:2005me} --- see Refs.~\cite{Liebling:2012fv, Visinelli:2021uve} for an extensive review.
A further reason for their relevance is their benign character in fully non-linear numerical relativity simulations, as they are regular solutions with ordinary matter sources that solve wave equations \cite{Bezares:2024btu}.
Their phenomenology has been studied from many perspectives, including their stability \cite{Lee:1988av, Gleiser:1988ih, DiGiovanni:2020ror, Siemonsen:2020hcg, Sanchis-Gual:2019ljs, Sanchis-Gual:2021phr, Brito:2025rld, Marks:2025jit, Marks:2025xxv}, 
behaviour and gravitational-wave emission in BS-BS \cite{Palenzuela:2006wp, Palenzuela:2017kcg, Bezares:2017mzk, Bezares:2018qwa, Bezares:2022obu, Helfer:2021brt, Sanchis-Gual:2022mkk, Siemonsen:2023age, Siemonsen:2023hko, Evstafyeva:2022bpr, Evstafyeva:2024qvp, Ge:2024itl,  Evstafyeva:2026juq} and
BH-BS binaries \cite{Clough:2018exo, Cardoso:2022vpj, Zhong:2023xll, Marks:2026xvo, Ning:2026qxs}, their
quasi-normal modes (QNMs) \cite{Yoshida:1994xi, Macedo:2013jja}, ringdown \cite{Siemonsen:2024snb}, the scalar interaction force \cite{Damour:2025oys} and whether they could reside at galactic centers \cite{Torres:2000dw, Guzman:2005bs, Amaro-Seoane:2010pks}.
Notably, open questions remain: for example, it is still an open question whether BSs with stable light rings (LRs), i.e.~those most strongly resembling BHs, are dynamically viable \cite{Cunha:2017qtt, Cunha:2022gde, Marks:2025jpt, Staelens:2025wom,  Evstafyeva:2025mvx}.
Additionally, rotating BSs have quantized angular momentum, which limits their astrophysical relevance, and those sufficiently compact to develop an ergoregion are likely subject to an associated instability \cite{Friedman:1978ygc, Vilenkin:1978uc, Moschidis:2016zjy, Siemonsen:2025fne, Siemonsen:2025wib, Siemonsen:2025ucx}. 
In general, it is still unclear to what extent BSs can approach the Kerr solution in the high-compactness limit --- though Ref.~\cite{Siemonsen:2025wib} points out that the LR frequencies obtained in Ref.~\cite{Siemonsen:2024snb}, which approach the Kerr value, may provide a hint.\\

With the publication of the Event Horizon Telescope (EHT) images \cite{EventHorizonTelescope:2019dse, EventHorizonTelescope:2022wkp}, many works have investigated to what extent they rule out ECOs as the central accreting objects. In many cases, the first thing to check is whether the ECO can give rise to a (partial) central brightness depression when surrounded by an emitting source to be qualitatively consistent with the EHT images. This has been done for a vast range of ECO models, e.g.~spherically symmetric BSs \cite{Rosa:2022toh, Rosa:2025dzq, Rosa:2023qcv, Li:2025awg, Zeng:2025nmu, Zeng:2025fjg, Wang:2026lmb}, rotating BSs \cite{Vincent:2015xta, Vincent:2020dij}, gravastars \cite{Aimar:2025uia}, fuzzballs \cite{Bacchini:2021fig, Mayerson:2023wck} and others \cite{Gyulchev:2021dvt, Olmo:2021piq, Gao:2025vov, Boos:2025nzc}. Models that cannot source a sufficiently strong brightness depression (e.g.~due to low compactness) are likely incompatible with the EHT results.

The absence of a horizon poses an additional problem with respect to the accretion mechanism. Typically, accretion discs are modelled by particles that follow stable timelike circular orbits (TCOs) up until the \textit{innermost stable circular orbit} (ISCO), while slowly exchanging energy and angular momentum through viscosity and friction. Within the ISCO, matter then rapidly falls into the BH horizon, marking a region of reduced emission. In an ECO spacetime, the logical expectation is that accreting matter would experience a similar infall and accumulate in the center. Constraints on emission from the central region are stringent, especially for Sgr A$^*$ --- see e.g.~Refs.~\cite{Narayan:2008bv, Broderick:2009ph}. Hence, to make a compelling case one would need a mechanism to suppress this central emission.
Along those lines, it was observed that matter accreting onto a BS can be (temporarily) stalled at an areal radius $r_\Omega > 0$ where the angular velocity for circular timelike geodesics peaks, due to a suppresion of the magnetorotational instability  \cite{Olivares:2018abq}. Recently, this has been further investigated for BSs that are not ultracompact, but still have a peak in the angular velocity function \cite{Jaramillo:2026ygy}. 
Ref.~\cite{Saurabh:2026ukk} studies accretion onto a horizonless singularity (JMN-1 spacetime), and compares it to its Schwarzschild counterpart, finding that they look relatively similar.
Ref.~\cite{Delgado:2021jxd} provides a thorough characterization of equatorial TCOs in an axisymmetric ultracompact spacetime, and finds that their structure is largely determined by the LRs and their stability properties. Additionally, they compute the efficiency associated with converting gravitational energy into radiation by adiabatic accretion. These works provide interesting first steps towards an understanding of what a realistic accretion process onto an ECO can look like, which is essential if one wants to model how ECOs would be observed by an experiment like the EHT.

It should be noted that other arguments have been put forward for the BH nature of M87$^*$ and Sgr A$^*$ in particular, which complement the EHT results. For example, significant constraints on a thermal or reflective surface exist \cite{Broderick:2009ph, EventHorizonTelescope:2022xqj}.\\

Full GRMHD simulations, like those in Refs.~\cite{Olivares:2018abq, Jaramillo:2026ygy}, are computationally very expensive, meaning that the number of simulations of accreting ECOs remains small. A cheaper approach simplifies the accretion disc to an analytic prescription, and subsequently creates images using geodesic ray-tracing\footnote{Note that this second step is also required if one wants to obtain images from GRMHD simulations.}. Multiple ray-tracing codes have been developed, but the majority is geared towards Kerr, making their adaptation for ECOs challenging. \texttt{FOORT} \cite{Mayerson:2025foo, MayersonFOORTgithub} --- see Sec.~\ref{subsec: theory - ray tracing} and App.~\ref{app: FOORT} --- has been developed to address this issue, providing a flexible and modular framework that is easily extended to non-Kerr spacetimes.\\

Due to large astrophysical uncertainties associated with the accretion disc, the diffuse direct emission in the EHT images is unlikely to provide strong constraints on deviations from the Kerr metric. Contrary, the so-called \textit{photon rings} are expected to provide stronger tests of general relativity and the Kerr metric  \cite{Gralla:2020srx}.
Photon rings arise as higher-order images of the accreted matter due to strong gravitational lensing around compact objects and are intimately connected with bound null geodesics --- see Sec.~\ref{subsec: theory - photon rings}. Photon rings can leave a universal imprint on the visibility amplitude over long baselines \cite{Johnson:2019ljv}. The photon rings and its properties have hence been investigated for non-Kerr BHs over the past years, to provide concrete alternative hypotheses that can be tested against --- see e.g.~Refs.~\cite{Staelens:2023jgr, daSilva:2023jxa}.
Most compact ECOs (and even modified BHs) can source additional photon rings in images, in particular when more LRs are present \cite{Olmo:2021piq, Cunha:2017wao, Shaikh:2019itn}.\\

We underline that the EHT does not make an \textit{image} in the traditional sense: data obtained through very long baseline interferometry (VLBI) is sampled in the \textit{visibility plane}, which effectively provides a (sparsely sampled) Fourier transform of the image. Hence, the most powerful ways to test the Kerr metric would use features in this visibility plane. Ref.~\cite{Johnson:2019ljv} argues that in the visibility plane individual photon rings would dominate different ranges of the baseline. Along these lines, it has been suggested that ECOs with more photon rings could reveal themselves through additional modulations  
\cite{Gao:2024ksc, Wang:2025hzu, Gao:2025vov}, which could be detected by Earth-Moon baselines. The capabilities of an EHT-like experiment to resolve double photon rings have been assessed in Ref.~\cite{Carballo-Rubio:2023ekp}, though the authors use synthetic images of two thin, uniform rings.\\

In this work, we use \texttt{FOORT} and its built-in functions to investigate the interferometric signatures associated with the appearance of double photon rings in ray-traced images of ultracompact BS spacetimes. In the rotating case, we further interpret the differences between BSs and the Kerr BH. In Sec.~\ref{sec: theory}, we summarise the relevant theory. Sec.~\ref{sec: results} presents our main results: we recover the expected modulation in the case of spherical symmetry, and discuss the strong breaking of our assumptions for rotating BSs. We summarise and conclude in Sec.~\ref{sec: conclusions}. We work in units such that $c = G = \hbar = 1$.

\section{Theory}\label{sec: theory}

This section aims to summarise the background material needed to understand the results in Sec.~\ref{sec: results}. We review BSs in Sec.~\ref{subsec: theory - boson stars}, after which we explain how LRs source photon rings in Secs.~\ref{subsec: theory - light rings} and \ref{subsec: theory - photon rings}. We briefly summarise the structure of TCOs in Sec.~\ref{subsec: theory - timelike circular orbits}. Finally, we describe our ray-tracing set-up in Sec.~\ref{subsec: theory - ray tracing}, and outline how we obtain the corresponding visibility amplitudes in Sec.~\ref{subsec: theory - visibility amplitudes}.

\subsection{Boson stars}
\label{subsec: theory - boson stars}

Boson stars describe a solitonic, self-gravitating configuration of a complex scalar field $\varphi$. They are solutions to the Einstein-Klein-Gordon equations, derived from the minimally coupled action
\begin{equation}
    \mathcal{S} = \int \du ^4x\frac{\sqrt{-g}}{2}\left\{\frac{R}{8 \pi}  -\left[
  g^{\mu\nu}\nabla_{\mu}\bar{\varphi}\,\nabla_{\nu}\varphi
  +V(|\varphi|^2)\right] \right\}, \label{eq:action}
\end{equation}
where $R$ is the Ricci scalar, $g_{\mu\nu}$ is the spacetime metric, and an overbar denotes complex conjugation. Having a complex scalar field, rather than a real one, allows for solutions with a stationary metric and energy-momentum tensor \cite{Kaup:1968zz}.
In this work, we restrict ourselves to so-called \textit{solitonic} BSs, where the potential takes the form
\begin{equation} \label{eq:solitonic}
    V(|\varphi|^2)= \mu^2 |\varphi|^2 \left(
  1-2\frac{|\varphi|^2}{\sigma_0^2}
  \right)^2.
\end{equation}
Here, $\sigma_0$ is a parameter that, when taken to be sufficiently small, can give rise to highly compact solutions. These so-called \textit{thin-shell} models have an energy density dominated by a sharp peak in the gradient term at finite radius~\cite{Collodel:2022jly, Boskovic:2021nfs}.

\begin{figure}
    \centering
    \includegraphics[width=0.9\linewidth]{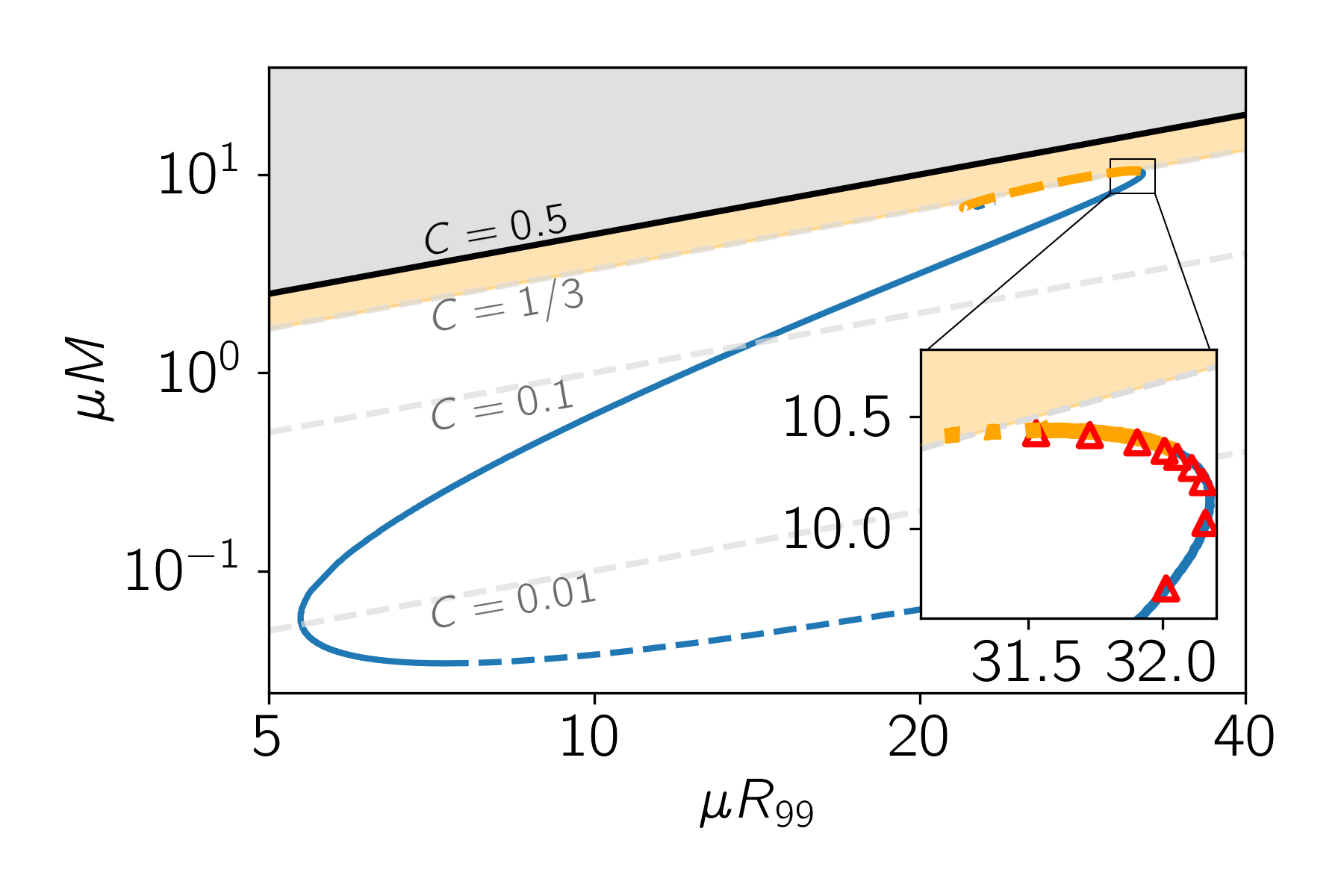}
    \caption{Mass $M$ v. radius\footnote{The radius of a BS is not well defined. Here, we adopt the commonly used areal radius containing 99\% of the mass.} $R_{99}$ for the family of spherically symmetric solitonic BSs with $\sigma_0 = 0.06$. Lines of constant compactness $C:= M/R_{99}$ are shown as well, with $C = 0.5$ corresponding to the BH limit, and $C = 1/3$ corresponding to the minimum compactness for an object to have the Schwarzschild unstable LR. Solid (dashed) sections correspond to models that are perturbatively (un)stable, and yellow sections correspond to ultracompact models, i.e.~models that have LRs. The red triangles correspond to the set of models used in this work.}
    \label{fig: m vs r s06}
\end{figure}

In the case of spherical symmetry, the line element in polar-areal coordinates takes the form

\begin{equation}\label{eq: sph symm metric}
  \du s^2 = -\alpha(r)^2\du t^2 + X(r)^2\du r^2 + r^2\du \Omega^2,
\end{equation}
and we look for solutions under a harmonic Ansatz for the scalar field $\varphi = A(r)e^{\iu\omega t}$. Upon fixing the potential, i.e.~choosing\footnote{In principle, we should also specify the scalar mass $\mu$. However, this simply rescales the BS solutions. As an example, for the spherically symmetric BSs considered here to have a mass on the order of $\sim 10^6\, M_\odot$, the scalar field mass should be on the order of $\sim 10^{-16}$ eV.} $\sigma_0$, the family of solutions can be parametrized by e.g.~the central scalar amplitude $A_0$, upon which the equations become an Eigenvalue problem for the frequency $\omega$ of the scalar field\footnote{Strictly speaking, $A_0$ is no longer a good parametrization for sufficiently small $\sigma_0$, as the family becomes multivalued in $A_0$. A more robust parametrization is given by e.g.~the central lapse $\alpha(r=0)$, but even with $A_0$ the BSs can be easily distinguished by e.g.~their mass $M$.}. We use an adapted version with quadruple precision of the two-way shooting code described in \cite{Evstafyeva:2023kfg}. In spherical symmetry, we focus on the family of solutions corresponding to  $\sigma_0 = 0.06$, shown in Figure \ref{fig: m vs r s06}. 

In the case of rotating BSs, we follow \cite{Siemonsen:2020hcg} and write the metric in the form
\begin{align}
ds^2 = - f\, dt^2
+ l f^{-1} \Big[&
g \left( dr^2 + r^2 d\theta^2 \right) \nonumber\\
&+ r^2 \sin^2\theta \left(d\phi - \Omega r^{-1} dt \right)^2
\Big]\,, \label{eq: rot metric}
\end{align}
where all metric functions depend only on $r$ and $\theta$. The scalar field now takes the form $\varphi = A(r, \theta)e^{\iu\omega t + i m \phi}$, where $m$ is an azimuthal index that quantises the angular momentum of the BS. Solutions to the EKG equations can be obtained using e.g.~a relaxation method \cite{Siemonsen:2020hcg}. 
We consider three models with $m=3$ used in Ref.~\cite{Siemonsen:2024snb} (and the Supplemental Material), whose properties\footnote{Note that our conventions differ from Ref.~\cite{Siemonsen:2024snb}, such that $\sigma_0 = \sigma_{\rm ref} \sqrt{2}$, with $\sigma_{\rm ref}$ the value quoted in Ref.~\cite{Siemonsen:2024snb}} we summarise in Table \ref{tab: BS models}. It should be noted that these models are likely unstable when evolved outside of axisymmetry \cite{Siemonsen:2020hcg, Sanchis-Gual:2019ljs}.
Additionally, we consider one model with $m=1$ that has lower angular momentum, and is also somewhat less compact. 
Note in particular that all of the models have $J/M^2 < 1$, which facilitates a comparison with Kerr\footnote{There is no analogue to the Birkhoff theorem beyond spherical symmetry that constrains the metric outside of the BS --- i.e., where the scalar field amplitude is negligible --- to be that of Kerr with the same angular momentum. Therefore, the comparison is not as straightforward as in the spherically symmetric case.}.

\begin{table}[]
    \centering
    \begin{ruledtabular}
    \begin{tabular}{lccccc}
    Model & $\sigma_0$ & $\mu M$ & $C$ & $J/ M^2$  & $n_{\rm LR}$ \\ \hline
    \texttt{A0437} & $0.06$ & 10.2 & 0.31 & 0 & 0\\
    \texttt{A044} & $0.06$ & 10.4 & 0.33 & 0 & 2\\
    \hline
    Model & $\sigma_0 / \sqrt2$ & $\mu M$ & $C$ & $J/ M^2$  & $m$ \\ \hline
    \texttt{C33} & 0.05 & 6.12 & 0.33 & 0.68 & 1 \\
    \texttt{C38} & 0.035 & 18.7 & 0.38 & 0.95 & 3 \\
    \texttt{C39} & 0.035 & 19.5 & 0.39 & 0.94 & 3 \\
    \texttt{C41} & 0.035 & 20.7 & 0.41 & 0.92 & 3\\
    \end{tabular}
    \end{ruledtabular}
    \caption{Summary of relevant properties for the reference spherically symmetric and rotating BS metrics used in this work. We give the solitonic parameter $\sigma_0$ (the $\sqrt{2}$ facilitates comparison with \cite{Siemonsen:2024snb, Evstafyeva:2025mvx}), the mass $M$, the compactness $C$, the dimensionless angular momentum $J / M^2$. For the spherically symmetric models, we give the number of LRs $n_{\rm LR}$, indicating whether the model is ultracompact, and for the rotating models we give the azimuthal index $m$.}
    \label{tab: BS models}
\end{table}

\subsection{Light rings}
\label{subsec: theory - light rings}

Objects that are sufficiently compact are able to support LRs: circular null geodesics in the equatorial plane. In suitable coordinates, they are located at fixed radii that correspond to extrema of the effective geodesic potential --- see e.g.~Ref.~\cite{Cunha:2017qtt}. For the metric \eqref{eq: rot metric}, suitable potentials are given by
\begin{equation}\label{eq: potentials axi}
    H_\pm = \frac{\Omega}{r} \pm \frac{f}{r\sin\theta\sqrt{l}}\,,
\end{equation}
where the plus (minus) sign corresponds to prograde (retrograde) motion. The stability of the LRs is then inferred from the second derivative of the potential: a prograde (retrograde) LR is stable if $\partial_r^2 H_+$ ($-\partial_r^2 H_-$) is positive, and unstable if it is negative.
As is well known, the Schwarzschild spacetime has one unstable LR at areal radius $r = 3M$, and the Kerr spacetime contains two unstable LRs, one corotating and one counterrotating, at areal radii that depend on the angular momentum. More so, the Kerr spacetime contains a \textit{photon shell}, a compact region between the two LRs where non-equatorial bound geodesics are allowed --- see e.g.~Ref.~\cite{Johnson:2019ljv} for an illustration. In particular, the photon shell also contains a polar LR, i.e.~a bound null geodesic that crosses the rotation axis.\\

Exotic compact objects (ECOs) need not have LRs, but once they become \textit{ultracompact}, LRs must appear \textit{in pairs} \cite{Cunha:2017qtt}, of which one is stable and one is unstable. It has been argued that the stable LR in particular could source a dynamical instability due to the slow decay of massless modes \cite{Keir:2014oka}, which has been posited as a strong argument against ultracompact BH mimickers: indeed, if this mechanism were efficient in destroying any object with a stable LR, all ultracompact objects would necessarily be BHs, as an event horizon is needed to generate unstable LRs in isolation. The debate on whether this is a genuine instability, and/or whether it is efficient, is still ongoing
\cite{Cardoso:2014sna, Marks:2025jpt, Staelens:2025wom, Cunha:2022gde, Evstafyeva:2025mvx, Guo:2024cts, Redondo-Yuste:2025hlv, Benomio:2024lev, Cunha:2025oeu}. In spherical symmetry, we focus on one of the ultracompact BSs from Ref.~\cite{Marks:2025jpt} that was shown to remain stable on long numerical timescales, specifically the \texttt{S06A044} model --- henceforth simply denoted as \texttt{A044} --- and models that are nearby in the parameter space; see Fig.~\ref{fig: m vs r s06}.

The rotating models in Table \ref{tab: BS models} are all ultracompact. The \texttt{C33} model has been the subject of recent LR instability investigations (model 2 in \cite{Cunha:2022gde}, model \texttt{S005} in \cite{Evstafyeva:2025mvx}) but has been shown to be stable on long numerical timescales \cite{Evstafyeva:2025mvx}. 
Additionally, it should be noted that all these models have an unstable and a stable counter-rotating LR, but lack co-rotating counterparts. Therefore, they do not have \textit{complete photon shells} in a way that a Kerr BH does: of course, the analogy does not quite hold, as the geodesic equations for the BS models are not separable in the $r$ and $\theta$ coordinates, like they are for Kerr.
As noted in the Supp.~Mat.~of \cite{Siemonsen:2024snb}, models \texttt{C39} and \texttt{C41} do have (one or more) polar null orbits.

\subsection{Photon rings}
\label{subsec: theory - photon rings}

Unstable light rings, and more generally any unstable bound null geodesics, are believed to source the so-called \textit{photon rings} that are expected to be present in (high-resolution) images of accreting BHs. They are generated by light that travels on geodesics that experience strong gravitational lensing, approximating these unstable bound null geodesics for a while, before escaping to infinity. More precisely, in theory one can expect an infinite self-similar sequence of photon rings, indexed by the number of times the geodesic passes through the equatorial plane and the accretion disc --- see e.g.~Ref.~\cite{Lupsasca:2024wkp} for a pedagogical introduction, as well as references therein. In the case of the Kerr metric, the shape of the photon rings for an observer at inclination $\theta_o$ with respect to the rotation axis is determined only by the subset of the photon shell where bound geodesics actually reach $\theta_o$ --- see e.g.~Fig.~2 in Ref.~\cite{Johnson:2019ljv} and Eq.~(4.13) in Ref.~\cite{Staelens:2023jgr}. In particular, an observer on the rotation axis sees a perfectly circular critical curve dictated by the properties of the unique polar LR.\\

In a bit more detail; the dynamics of these nearly bound geodesics are governed by the Lyapunov exponent $\gamma_L$ according to \cite{Bozza:2002zj, Johnson:2019ljv}
\begin{equation}
    \label{eq: Lyapunov}
    \delta r_n \approx e^{\gamma_L\, n} \delta r_0\,.
\end{equation}
In this expression, $\delta r_n = r_n - r^u_{\rm LR}$ is the radial growth of the nearly-bound geodesic after $n$ half-orbits --- we refer the reader to App.~\ref{app: Lyapunov} for more details.
This exponent is well studied for the Kerr BH (see e.g.~Refs.~\cite{Johnson:2019ljv, Lupsasca:2024wkp}), but also for BHs beyond general relativity (e.g.~Ref.~\cite{Staelens:2023jgr, Deich:2023oox}). Successive photon rings are expected to decrease exponentially in width and flux, at a rate set by $\gamma_L$, making this quantity potentially observable. Additionally, these photon rings rapidly approach the \textit{critical curve}, whose shape is accurately predicted from theory and could form a strong test of the Kerr metric \cite{Johnson:2019ljv, Gralla:2020srx}.\\

Ref.~\cite{Cardoso:2008bp} derives an expression for the Lyapunov exponent in a generic spherically symmetric spacetime, although the convention for the definition of the exponent differs. In this latter convention, there is a strong connection between the Lyapunov exponent and the damping timescale of quasi-normal modes (QNMs) \cite{Ferrari:1984zz, Schutz:1985km, Pedrotti:2025idg}.
We present the analogous formula for $\gamma_L$ in the imaging conventions in Sec.~\ref{subsec: results - Lyapunov} for the spherically symmetric metric \eqref{eq: sph symm metric}, as well as an expression for $\gamma_L$ associated with the unstable LR in rotating BS spacetimes.

Finally, we note that the presence of polar LRs in the \texttt{C39} and \texttt{C41} models
suggests that even an on-axis observer could see an infinite sequence of photon rings around these objects. Conversely, we expect this to not be the case for te \texttt{C33} and \texttt{C38} models.

\subsection{Timelike circular orbits}
\label{subsec: theory - timelike circular orbits}

In this section, we briefly address the structure of TCOs in the BS spacetimes, as this is relevant for the matter in the accretion disc.

It is well known that the Schwarzschild BH allows TCOs at radii larger than the unstable LR at $r=3M$, and that these are only stable outside of the ISCO at $r=6M$. The Kerr BH allows for a similar structure, although we now have to differentiate between prograde and retrograde TCOs. The orbital angular frequency of TCOs in a general axisymmetric, stationary spacetime can be calculated as
\begin{equation}\label{eq: orbital angular frequency}
    \Omega^\pm = \frac{-g'_{t\phi} \pm \sqrt{(g'_{t\phi})^2 - g_{tt}'g_{\phi\phi}'}}{g_{\phi\phi}'}\,,
\end{equation}
where primes denote derivatives with respect to the radial coordinate, and $t,\phi$ are the standard coordinates adapted to the symmetry. Circular timelike orbits only exist when the expression under the square root is positive.

\begin{figure}[t]
    \centering
    \includegraphics[width=\linewidth]{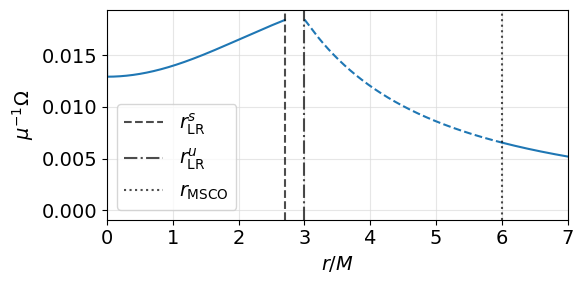}
    \caption{Orbital angular frequency for TCOs as function of the areal radius $r$, in the \texttt{A044} spacetime. Solid (dashed) segments correspond to (un)stable TCOs. The vertical lines indicate the LRs and the MSCO. The region between the two LRs does not allow TCOs --- see Ref.~\cite{Delgado:2021jxd}.}
    \label{fig: Omega sph BS}
\end{figure}

Figure~\ref{fig: Omega sph BS} shows the angular frequency of TCOs for the \texttt{A044} spacetime, as well as the locations of the LRs and the \textit{marginally stable circular orbit}\footnote{Defined in Ref.~\cite{Delgado:2021jxd} as \textit{the stable timelike CO with the smallest radius that is continuously connected to spatial infinity by a set of stable TCOs}.} (MSCO). In between the LRs, Ref.~\cite{Delgado:2021jxd} showed that no TCOs exist, and inside the stable LR, stable TCOs again exist all the way to the centre. This implies that, in principle, an inner accretion disc could form.

\begin{figure}[t]
    \centering
    \includegraphics[width=\linewidth]{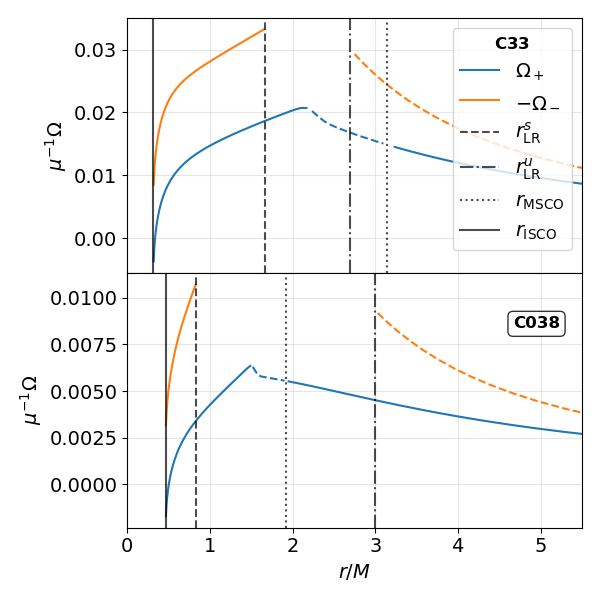}
    \caption{The orbital angular frequencies (\ref{eq: orbital angular frequency}) for the (\textit{top}) \texttt{C33} and (\textit{bottom}) \texttt{C38} models, as a function of the areal radius $r$. Solid (dashed) segments correspond to (un)stable TCOs. The vertical lines indicate the (retrograde) LRs and the ISCO and MSCO for co-rotating TCOs. The region between the two LRs does not allow retrograde TCOs.}
    \label{fig: radial structure both}
\end{figure}

Figure \ref{fig: radial structure both} shows the orbital angular frequency $\Omega^\pm$ for the \texttt{C33} and \texttt{C38} models.
 Given that both LRs are counter-rotating, Ref.~\cite{Delgado:2021jxd} only guarantees that retrograde TCOs cannot exist in between. Indeed, we find that prograde TCOs exist everywhere up to the ISCO, although a prograde (and retrograde) MSCO exists below which TCOs are unstable over a range of radii.

\begin{table}[]
    \centering
    \begin{ruledtabular}
    \begin{tabular}{lcccc}
        Model & $\mu r_{\rm MSCO}^+$ & $\mu r_{\rm LR}^{s}$ & $\mu r_{\rm LR}^{u}$ & $N_x$ \\ \hline
\texttt{C33} & 19 & 10.2 & 16.5 & 500 \\
\texttt{C38} & 36 & 15.5 & 56 & 1500 \\
\texttt{C39} & 38 & 14.5 & 58 & 500  \\
\texttt{C41} & 38 & 12.5 & 60 & 500 \\
\hline
\texttt{A044} & 62.4 & 28.1 & 31.2 & NA \\
    \end{tabular}
    \end{ruledtabular}
    \caption{Summary of relevant radii for the reference BS models used in this work --- see Table \ref{tab: BS models}. We provide \textit{approximate} values for the radial coordinates of the MSCO and (un)stable LR: see App.~\ref{app: FOORT} for a brief comment on the resolution of the BS metrics, related to the $N_x$ column.}
    \label{tab: BS models radii}
\end{table}

\subsection{Ray-tracing}
\label{subsec: theory - ray tracing}

The images in this work are obtained with \texttt{FOORT}, a Flexible Object-Oriented Ray Tracer that is publicly available on GitHub \cite{Mayerson:2025foo, MayersonFOORTgithub}. 
Ray-tracing essentially evolves the geodesic equation backwards in time from a far-away observer to predict how they would perceive the (environment of the) central compact object.
\texttt{FOORT} is easily modified and designed to handle non-Kerr metrics, and has been used to image a variety of spacetimes, including BSs and fuzzballs \cite{Staelens:2023jgr, Mayerson:2023wck, Bacchini:2021fig, Fernandes:2024ztk, Carballo-Rubio:2025zwz, Marks:2025jpt}; more details are provided in App.~\ref{app: FOORT}.\\

For the purpose of this work, we assume that there is some geometrically thin, optically thick accretion disc that is confined to the equatorial plane\footnote{The rotating BS metrics --- as well as the spherically symmetric ones --- are symmetric with respect to the plane $\theta= \frac\pi2$, making an equatorial disc natural.}. To model the intensity profile of the disc, we use the GLM model \cite{Gralla:2020srx} which has also been used in e.g.~Ref.~\cite{Rosa:2023qcv} to image BSs, and is implemented in \texttt{FOORT}. The GLM intensity profile is given by
\begin{equation} \label{eq: GLM profile}
    I_0(r|\rho, \gamma,\sigma) = \frac{\exp\left(-\frac12\left[\gamma + \operatorname{arcsinh}\left(\frac{r-\rho}{\sigma}\right)\right]^2\right)}{\sqrt{(r-\rho)^2+
    \sigma^2}}\,,
\end{equation}
where $\rho$ controls the peak of the emission, $\sigma$ is the decay width, and $\gamma$ determines the asymmetry between the in- and outside of the disc. Of course, this is a highly simplified emission source: more realistic sources would be geometrically thick (and possibly toroidal \cite{Meliani:2015zta, Vincent:2015xta, Vincent:2020dij}), time-dependent, and inclusive of GRMHD physics (potentially including more emission from within the plunging region \cite{Mummery:2024mrq}). The GLM model is generally considered to be quite informative, despite its simplicity, and is especially useful when compared to time-averaged, realistic images \cite{Gralla:2020srx, Johnson:2019ljv, Lupsasca:2024wkp}.

In the case of spherically symmetric BSs, we focus on the case where the emission peaks around the Schwarzschild ISCO, at a value $\rho = 6M$. The Birkhoff theorem guarantees\footnote{Strictly speaking, the scalar field outside $R_{99}$ never drops to zero, but rather follows an exponential decay. Hence, one can treat the spacetime sufficiently far outside $R_{99}$ as \textit{essentially} vacuum.} that our compact BSs have a MSCO at this radius.
We further take $\gamma = -2$ and $\sigma = M/4$ as in Ref.~\cite{Rosa:2023qcv}. We will refer to the above disc model as the \textit{MSCO model}. 
In the case of rotating BSs, we also choose the peak $\rho$ to be at the MSCO, but this radius now depends on the model and is listed (approximately) in Table \ref{tab: BS models radii}.

While the above disc model is a standard choice for BHs, accretion discs around ECOs could in principle take different forms, motivated by the absence of an event horizon and the existence of stable TCOs inside the MSCO --- see Figs.~\ref{fig: Omega sph BS} and \ref{fig: radial structure both}. 
To this end, we also consider an alternative disc model motivated by Refs.~\cite{Olivares:2018abq, Jaramillo:2026ygy}, where the intensity peaks at the radius $r_\Omega$, corresponding to the radius at which the angular velocity profile has an extremum. Details and results are given in Sec.~\ref{subsec: results - stalled disc}.
Finally, in Appendix \ref{app: central disc model} we briefly consider an accretion disc that persists all the way to the core of the BS, as was done in e.g.~Ref.~\cite{Rosa:2023qcv}.\\

The total brightness intensity for the geodesic, as perceived by the observer, is calculated as
\begin{equation}\label{eq: I tot}
    I_{\text{tot}} = \sum_{n=1}^{N_{\text{max}}} \zeta_n I_0(p^{(n)}) g(p^{(n)})^f\,.
\end{equation}
In this expression, $N_{\text{max}}$ is the number of times the geodesic passes through the equatorial plane\footnote{$N_{\text{max}}$ can also be set to a limiting value, thus avoiding higher order contributions to the intensity and mimicking optical thickness.} and $\zeta_n$ is a geometric fudge factor used to mimic the effects of an optically thick disc: $\zeta_1 = 1$ and $\zeta_{n>1} = \zeta \leq 1$.
$I_0(p^{(n)})$ is the local intensity at the $n^{\rm th}$ point of passing $p^{(n)}$, $g(p^{(n)})$ is the redshift factor and $f$ is the power of the latter, set to 3 by default\footnote{A power of 3 is a common choice in the literature, as $I_\nu / \nu^3$ is the invariant number of photons of frequency $\nu$ along a ray. With $f=3$, $I_{\text{tot}}$ is to be interpreted as a specific intensity under the assumption of a broadband source. 
Setting $f=4$ means that $I_{\text{tot}}$ is to be interpreted as bolometric \cite{Gralla:2020srx}.}. 

In addition to an emission profile, we also need a model for the source velocity of the matter in the accretion disc to determine the redshift factor $g(p)$. \texttt{FOORT} supports several options already established in the literature \cite{Cardenas-Avendano:2022csp}. In this work, we focus on matter in Keplerian orbits\footnote{Note that this leads to asymmetry in the intensity of the observed images due to Doppler beaming, even when the spacetime is spherically symmetric.} --- some details and alternatives are provided in App.~\ref{app: FOORT}.\\

Given that we have a free scale with the scalar mass $\mu$ we can rescale our images to any desired size: when needed, we will use this freedom to rescale our images to a fiducial angular width of 100 $\mu$as, comparable to the scale of the EHT images of M87${}^*$ and Sgr A${}^*$ \cite{EventHorizonTelescope:2019dse, EventHorizonTelescope:2022wkp}.

\subsection{Visibility amplitudes}
\label{subsec: theory - visibility amplitudes}

The EHT images are obtained through Very Long Baseline Interferometry (VLBI), a technique that fundamentally differs from e.g.~traditional optical telescopes. VLBI samples the complex \textit{visibility function}; essentially the Fourier transform of the intensity profile $I(\mathbf{x})$:
\begin{equation}
    V(\mathbf{u}) = \int I(\mathbf{x}) e^{-2\pi i\, \mathbf{x}\cdot \mathbf{u}} \: {\rm d}^2 \mathbf{x}\,.
\end{equation}
Here, $\mathbf{x}$ is a 2-dimensional vector representing dimensionless coordinates on the observer screen, and $\mathbf{u}$ are dimensionless baselines in units of the observer wavelength. We will typically be interested in the \textit{visibility amplitude} (VA), ignoring the details of the phase. In addition, we will plot the VA along a certain slice $S_\alpha := \{\mathbf{u} := (u_1, u_2)\in \mathbb{R}^2 |\, u_2 = u_1 \tan \alpha \}$. In principle, we should look at the full 2D visibility, but relevant features can typically be understood from a slice: where this is not the case, we show multiple slices of the VA. Plotting the VA along such a slice is implemented efficiently in \texttt{FOORT} by leveraging the \textit{projection-slice theorem}, which avoids calculating the full 2D Fourier transform.\\

Small baselines capture global properties of the intensity profile, whereas long baselines correspond to high-frequency details. Photon rings are expected to have a small width compared to their diameter $d$, causing them to leave imprints on the visibility function over long baselines \cite{Johnson:2019ljv}. In the case of an infinitesimally thin, uniform and circular disc, i.e.~$I(\mathbf{x}) = \frac{I_0}{\pi d} \delta\left(|\mathbf{x}| - \frac{d}{2}\right)$, the visibility amplitude takes the form
\begin{equation}
    V(u) = I_0 J_0\left(\pi d u\right)\,,
\end{equation}
where $J_0$ is the 0$^{\rm th}$ Bessel function of the first kind. For large baselines $u$, this is well approximated by
\begin{equation} \label{eq: visamp ring approx}
    V(u) \propto \frac{\cos\left(\pi d u - \pi /4\right)}{\sqrt{du}}\,. 
\end{equation}

This means that the photon ring leaves an imprint at the characteristic frequency $d/2$, inside an envelope that falls as $u^{-1/2}$. As explained in Sec.~\ref{subsec: theory - light rings}, subsequent photon rings are expected to decrease exponentially in width, i.e.~$w_{n+1} / w_n \approx e^{-\gamma_L}$. The $n^{\rm th}$ photon ring is expected to dominate the signal in the regime \cite{Johnson:2019ljv}
\begin{equation}
\label{eq: baseline dominate}
    \frac{1}{w_{n-1}} \ll u \ll \frac{1}{w_n}\,.
\end{equation}

It is well known that for horizonless ultracompact objects, and some modified BHs, images can contain more photon rings than is the case for Kerr \cite{Aimar:2025uia, Vincent:2015xta, Wang:2025hzu, Gao:2024ksc, Gao:2025vov, Carballo-Rubio:2023ekp, Gyulchev:2021dvt, Olmo:2021piq, Li:2025awg, Rosa:2023qcv, Rosa:2025dzq, Rosa:2022toh, Wang:2025hzu, Zeng:2025nmu}. In particular, photon rings can have counterparts inside the critical curve, which is not possible for Kerr BHs. In the case of spherical BS, two sequences of rings exist, both decreasing exponentially in width. This means that, in light of Eq.~\eqref{eq: baseline dominate}, at a given baseline, we expect two rings with comparable widths to dominate the VA together. Assume now that these photon rings are concentric and have comparable radii $d_1 \approx d_2 \approx \overline{d} := \frac{d_1 + d_2}{2}$. Using Eq.~\eqref{eq: visamp ring approx}, we then expect the VA to be of the form
\begin{equation} \label{eq: visamp ring approx 2}
    V(u) \propto \frac{\cos\left(\pi \overline{d} u - \pi /4\right) \cos\left(\pi \Delta d u / 2\right)}{\sqrt{\overline{d}u}}\,,
\end{equation}
i.e.~the total VA is similar to that of a single photon ring, but modulated at a low frequency $\Delta d := \frac{d_1 - d_2}{4}$. 

With the images having a fiducial width of $100$ $\mu$as, a difference in angular radius $\Delta \beta$ between the photon rings should correspond to a modulation with period
\begin{equation}\label{eq: T vs delta beta}
    \frac{T}{100\, G\lambda} \approx 4.1 \left(\frac{\Delta \beta}{\mu{\rm as}}\right)^{-1}
\end{equation}
in our VAs.

The above argument suggests that, at least in an idealised scenario, a beating envelope over long baselines should be a generic feature of a BH mimicker \cite{Wang:2025hzu}. Throughout the paper, we will focus on the first-order photon rings, i.e.~those generated by geodesics that pass through the accretion disc twice. Higher-order photon rings are expected to manifest themselves on longer baselines, making them even more challenging to detect. On the other hand, they are expected to be even more circular than first-order photon rings, making the above approximations likely more valid.

\section{Results} \label{sec: results}

\begin{figure*}[ht]
    \centering
    \includegraphics[width=0.8\linewidth]{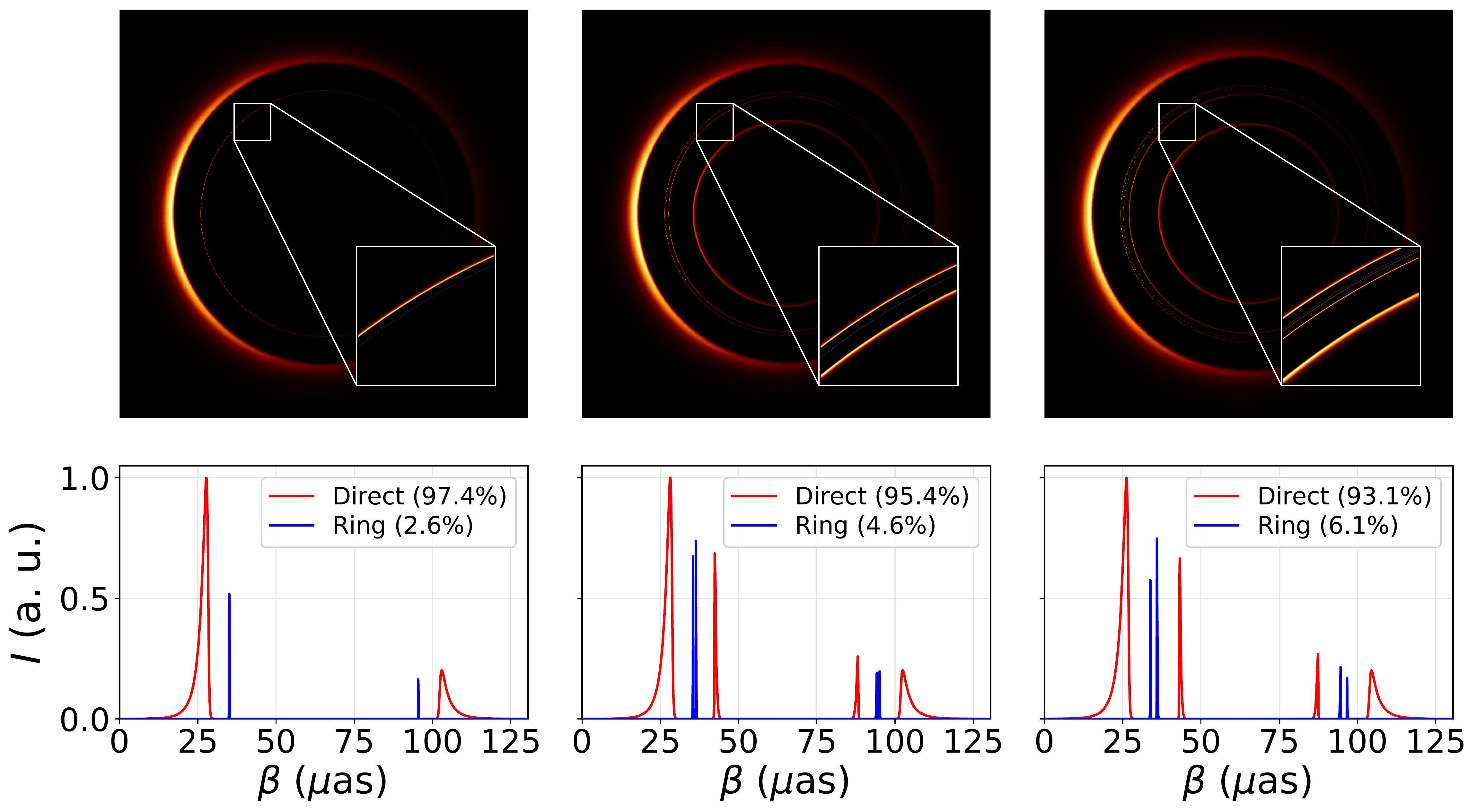}
    \caption{(\textit{top}) Ray-traced images, as seen by an equatorial observer and (\textit{bottom}) a slice through the intensity profile (at an angle of $40^\circ$). From left to right, we have a Schwarzschild BH, the \texttt{A0437} and the \texttt{A044} spherically symmetric BSs --- see Table \ref{tab: BS models}.
    The accretion disc emission follows the GLM profile \eqref{eq: GLM profile} with $\rho = 6M$, $\gamma = -2$,  $\sigma = M/4$, and assumes MSCO-model for the fluid velocity. The insets in the images zoom in on the ring structure.
    The intensity profiles in the bottom row are separated into direct emission and photon rings, with the percentages indicating the area under the curves.}
    \label{fig: Images and lineouts}
\end{figure*}

This section present the results of our analysis. In Sec.~\ref{subsec: results - spherical symmetry}, we find that the expected modulation is very clearly recovered for spherically symmetric, ultracompact BS where the observer is close to the equatorial plane. Interestingly, we find that a modulation persists over a small range outside the ultracompact regime, despite the absence of LRs. However, we also show that the modulations are suppressed when the observer inclination as a result of reduced symmetry. In Sec.~\ref{subsec: results - rotating BS} we present the results for some rotating BSs, and discuss how the comparison with Kerr is less straightforward. We show that the absence of modulations is not surprising given the complex photon ring structure, and interpret the latter as a result of the absent corotating LR. To make a connection with recent work \cite{Jaramillo:2026ygy}, we briefly consider a \textit{stalled} disc model in Sec.~\ref{subsec: results - stalled disc}. Finally, we present formulas for the Lyapunov exponent in Sec.~\ref{subsec: results - Lyapunov}.

\subsection{Spherical symmetry}
\label{subsec: results - spherical symmetry}

We compare the images and VAs of a sequence of BS models (Fig.~\ref{fig: m vs r s06}) with those of a Schwarzschild BH of equal mass $M$. The sequence spans both ultracompact and less compact models. 

We start by examining the images: Figure \ref{fig: Images and lineouts} shows the ray-traced images, as well as slices of the intensity profile, for two representative BS models alongside the Schwarzschild BH. In all cases, the accretion disc emission is modelled using the MSCO-model\footnote{We checked that the width of the disc only starts to affect the results when $\sigma \gtrsim M$, when the overall intensity distribution becomes diffuse.} described in Sec.~\ref{subsec: theory - ray tracing}. 
All spacetimes are scaled to the same physical mass $M$, and the image width is set to a fiducial $100\,\mu\mathrm{as}$, comparable to the angular scale of the EHT images of M87${}^*$ and Sgr A${}^*$ \cite{EventHorizonTelescope:2019dse, EventHorizonTelescope:2022wkp}.

\begin{figure}[t]
    \centering
    \includegraphics[width=1\linewidth]{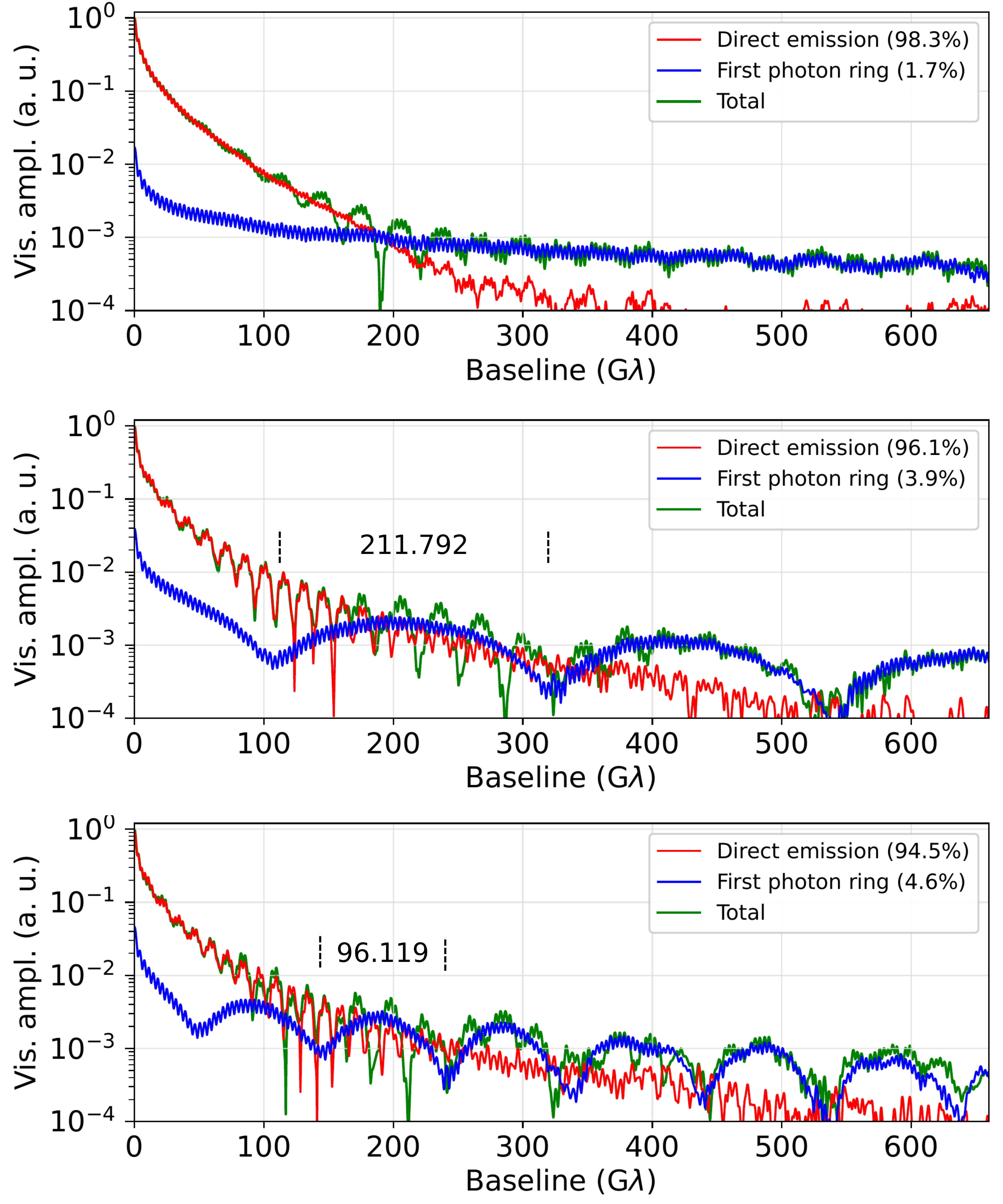}
    \caption{Normalised visibility amplitude $|V(u)|$ along a slice at $40^\circ$ as a function of baseline $u$ for the images in Fig.~\ref{fig: Images and lineouts}: (\textit{top}) a Schwarzschild BH and the (\textit{middle}) \texttt{A0437} and (\textit{bottom}) \texttt{A044} BSs. In each panel, the total visibility amplitude (green) is decomposed into  contributions from direct emission (red) and the first photon ring (blue), with percentages indicating the contribution to the total intensity integrated over the whole image. Estimates for modulation periods are indicated.}
    \label{fig: visamps sph symm}
\end{figure}

The Schwarzschild image exhibits the expected structure: a bright primary emission ring surrounding a central brightness shadow, with a narrow photon ring appearing in between. In contrast, the BS images display additional rings arising from the absence of an event horizon. Most notably, a small bright inner ring appears for each BS, representing a lensed image of the accretion disc emission that passes through the BS interior and re-emerges on the far side; a feature that is entirely absent in the BH case, where such geodesics are obstructed by the event horizon.

The lineout profiles (lower panels of Fig.~\ref{fig: Images and lineouts}) highlight the key structural difference between the BS and BH cases. For the BS models, the rings can be seen to produce nearby peaks in the intensity: the separation between the paired peaks increases as the BS becomes more compact. In the other direction, decreasing the compactness would eventually see the photon rings merge, once the spacetime no longer contains geodesics that are sufficiently lensed. Note, however, that the \texttt{A0437} model does not contain LRs, but still displays photon rings. In general, non-ultracompact models can in principle source a finite number of photon rings: it is only when the LRs appear that an infinite sequence can be produced.\\

\begin{figure}[t]
    \centering
    \includegraphics[width=1\linewidth]{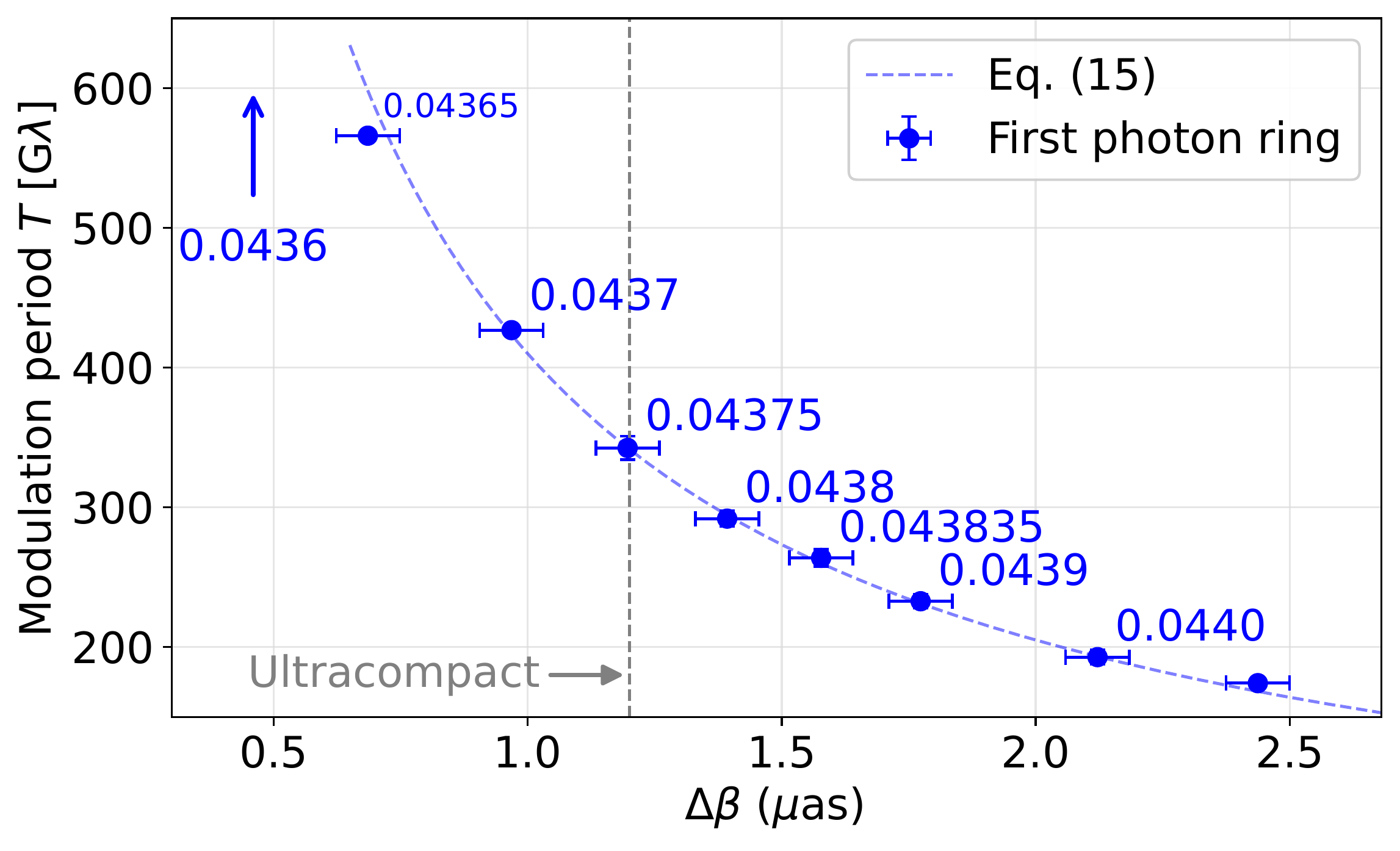}
    \caption{Estimated period of the low-frequency modulation in the visibility amplitude vs. the estimated separation of the photon rings in the ray-traced images, for the sequence of BS models shown in Fig.~\ref{fig: m vs r s06}.
    The annotations correspond to the central scalar amplitude of the different models, and the dashed vertical line indicates when the models become ultracompact. The blue dashed line shows the inverse proportionality relation \eqref{eq: T vs delta beta}, underlining that the modulation is indeed attributed to the small separation of the photon rings. The arrow indicates a model for which we could not robustly extract a modulation in the range of baselines considered, despite having distinct photon rings. Horizontal errorbars represent the resolution of the images, and (small) vertical errorbars represent the standard deviation in case multiple estimates for the period can be extracted.
    }
    \label{fig: T vs delta B}
\end{figure}

We now turn to Fourier space, to search for interferometric signatures of the ultracompact BS models. Figure ~\ref{fig: visamps sph symm} shows the normalised visibility amplitude $|V(u)|$ as a function of the baseline length $u$ for the models in Fig.~\ref{fig: Images and lineouts}. 
In each panel, the total visibility amplitude is decomposed into contributions from the direct emission and the first photon ring: this is done by tracking the number of passes the geodesics make through the equatorial plane.
In all cases, the direct emission dominates at short baselines and decays. As the baseline increases, the photon ring contribution starts to dominate the total signal --- this is exactly the behaviour described in Ref.~\cite{Johnson:2019ljv}. For the Schwarzschild BH, the (slow) decay persists, but for the BSs we can see that low-frequency modulations appear in the photon ring contribution (and hence the total). We briefly discuss the numerical accuracy of our VAs in App.~\ref{app: convergence}, where we find that they are trustworthy above $\gtrsim 10^{-4}$, making these modulations robust.

To verify that these modulations are those predicted by Eq.~\eqref{eq: visamp ring approx 2}, we estimate the modulation period by extracting relative minima in the VA\footnote{We preprocess the VA with a Gaussian filter to recover the minima robustly.}: the separation between two successive minima then represents half a period. This is illustrated in Fig.~\ref{fig: T vs delta B}, which shows that the extracted modulation periods follow the expected inverse behaviour \eqref{eq: T vs delta beta} with respect to the separation between the photon rings. Fig.~\ref{fig: T vs delta B} also highlights that we do detect these low-frequency modulations for BS models that are just outside of the ultracompact regime, as the lensing is still sufficiently strong to have geodesics pass through the accretion disc multiple times, despite the absence of any LRs. Note that the baseline scales involved here, at $\mathcal{O}(100\, {\rm G}\lambda)$, are well beyond the current capabilities of the EHT, and would likely require Moon- or space-based telescopes to join the array\footnote{Note that the proposed BHEX mission \cite{Johnson:2024ttr} is targeted to provide additional baselines around $\sim 10\, {\rm G}\lambda$, which are therefore likely too small.}.\\

\begin{figure*}
    \centering
    \includegraphics[width=0.7\linewidth]{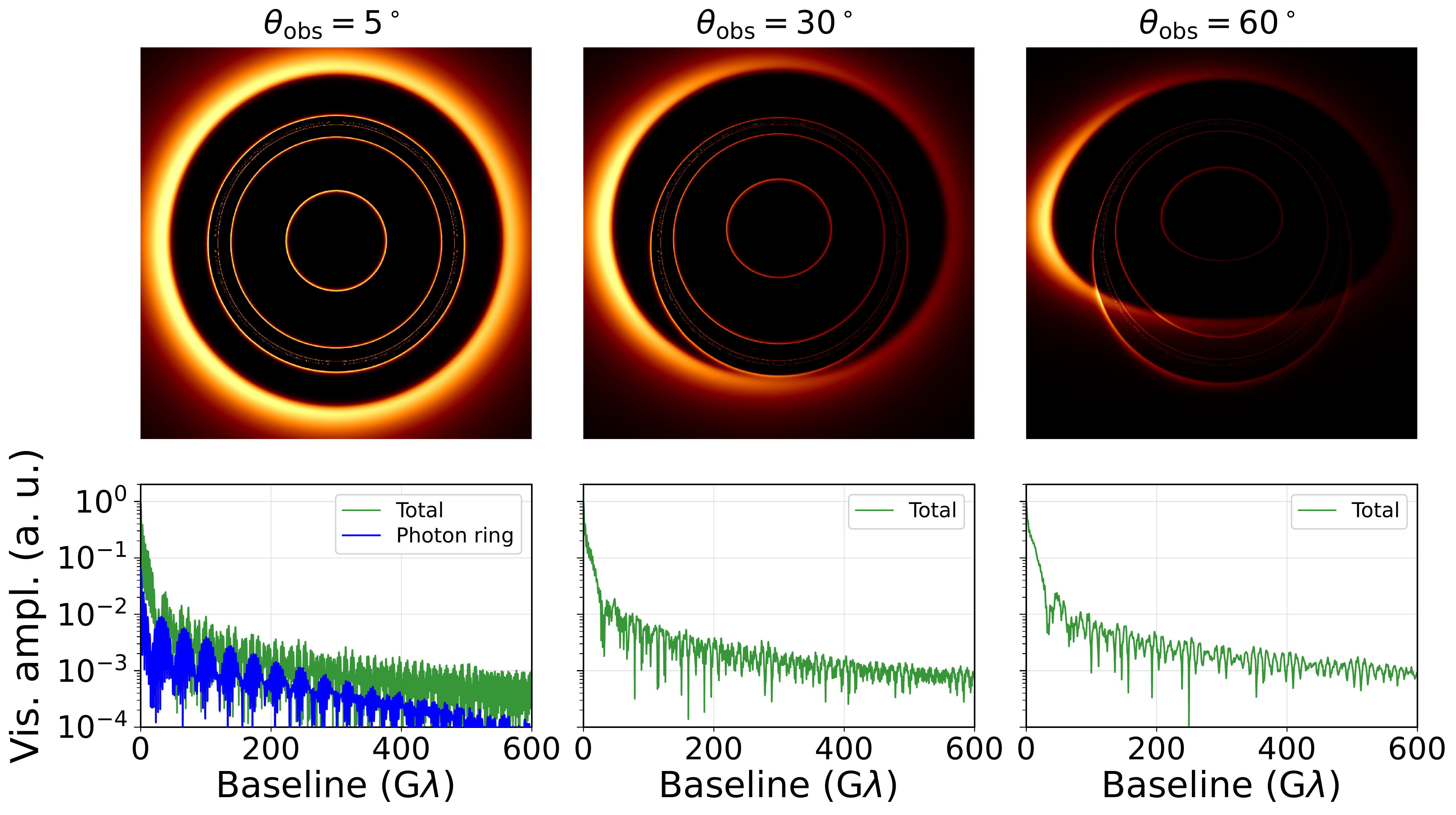}
    \caption{(\textit{top}) Ray-traced images and (\textit{bottom}) visibility amplitudes for the \texttt{A044} BS as seen by observers at inclinations of $5^\circ$, $30^\circ$, and $60^\circ$. The images are rescaled to a fiducial angular width of 100 $\mu$as. The total visibility amplitude is shown, and for the observer at $\theta_o = 5^\circ$, the visibility amplitude associated with the first-order photon rings is also shown.}
    \label{fig: modulation angles}
\end{figure*}

Thus far, we have assumed an observer located in the equatorial plane, i.e.~at an inclination $\theta_o = 90^\circ$. This is a favourable setup, as the spacetime is symmetric with respect to the equatorial plane: the result is that the photon rings appear as concentric rings, with the only asymmetry being due to Doppler beaming of the rotating disc: this is in line with some of the assumptions in Sec.~\ref{subsec: theory - visibility amplitudes}. Moving away from the equatorial plane, the photon rings will generally no longer be concentric or circular\footnote{These deviations are expected to disappear for higher-order photon rings, as they rapidly approach the critical curve. For a spacetime with equatorial symmetry, this critical curve has a reflection symmetry (see e.g.~Ref.~\cite{Cunha:2018uzc}), and for spherical symmetry it is circular.}, which could break the assumptions going into Eq.~\eqref{eq: visamp ring approx 2}. We consider three off-equatorial observers in Fig.~\ref{fig: modulation angles}, where the observer at $\theta_o = 5^\circ$ is nearly on-axis\footnote{Due to the polar singularity of the metric, the image for an observer that is exactly on axis suffers from numerical artifacts.}.

Indeed, the photon rings for the intermediate observers at $\theta_o = 30^\circ, 60^\circ$ are not concentric circles. Fig.~\ref{fig: modulation angles} also shows the total VAs: while there is some evidence for a low-frequency modulation for the observer at $60^\circ$, it is not as clear as Fig.~\ref{fig: Images and lineouts}. The absence of any obvious modulation for the observer at $\theta_o = 30^\circ$ is even more surprising, as the photon rings are more circular. At the moment, \texttt{FOORT} is not set up to disentangle direct emission and photon rings in the case of overlap, meaning that investigating their separate contributions, like we did in Fig.~\ref{fig: Images and lineouts} is not straightforward. We defer this, as well as a more in-depth investigation of why the modulations are absent, to future work.

For the observer at $\theta_o = 5^\circ$, the image again consists of concentric rings due to the axisymmetry of the spacetime. For this set-up, we can again separate the total VA into individual components, and we see that the contribution of the photon ring again shows the expected modulations (with a shorter period compared to Fig.~\ref{fig: Images and lineouts}, as the photon rings are separated more). However, in this case it does not dominate the total signal as strongly as it did in Fig.~\ref{fig: Images and lineouts}. A potential explanation here could be that the innermost lensed image of the direct emission now has a width comparable to that of the photon rings, breaking the assumption that two rings dominate the VA at certain baselines (see Eq.\eqref{eq: baseline dominate}). We also note that the photon ring contribution now decays faster, with respect to the total.\\

We conclude that the low-frequency modulation forms a clear and dominant part of the total VA for (near)-equatorial observers, whereas some of the assumptions that go into Eq.~\eqref{eq: visamp ring approx 2} may be broken for observers that are sufficiently far away from the equator. We hence conclude that, at least for the first-order photon rings, a strong, low-frequency modulation may not be present in generic images of ECOs. We reiterate that this picture could again change for the interference between higher-order photon rings, which are further circularized, but contribute to the signal at even larger baselines.

\subsection{Rotating Boson Stars}
\label{subsec: results - rotating BS}

\begin{figure*}
    \centering
    \includegraphics[width=0.7\linewidth]{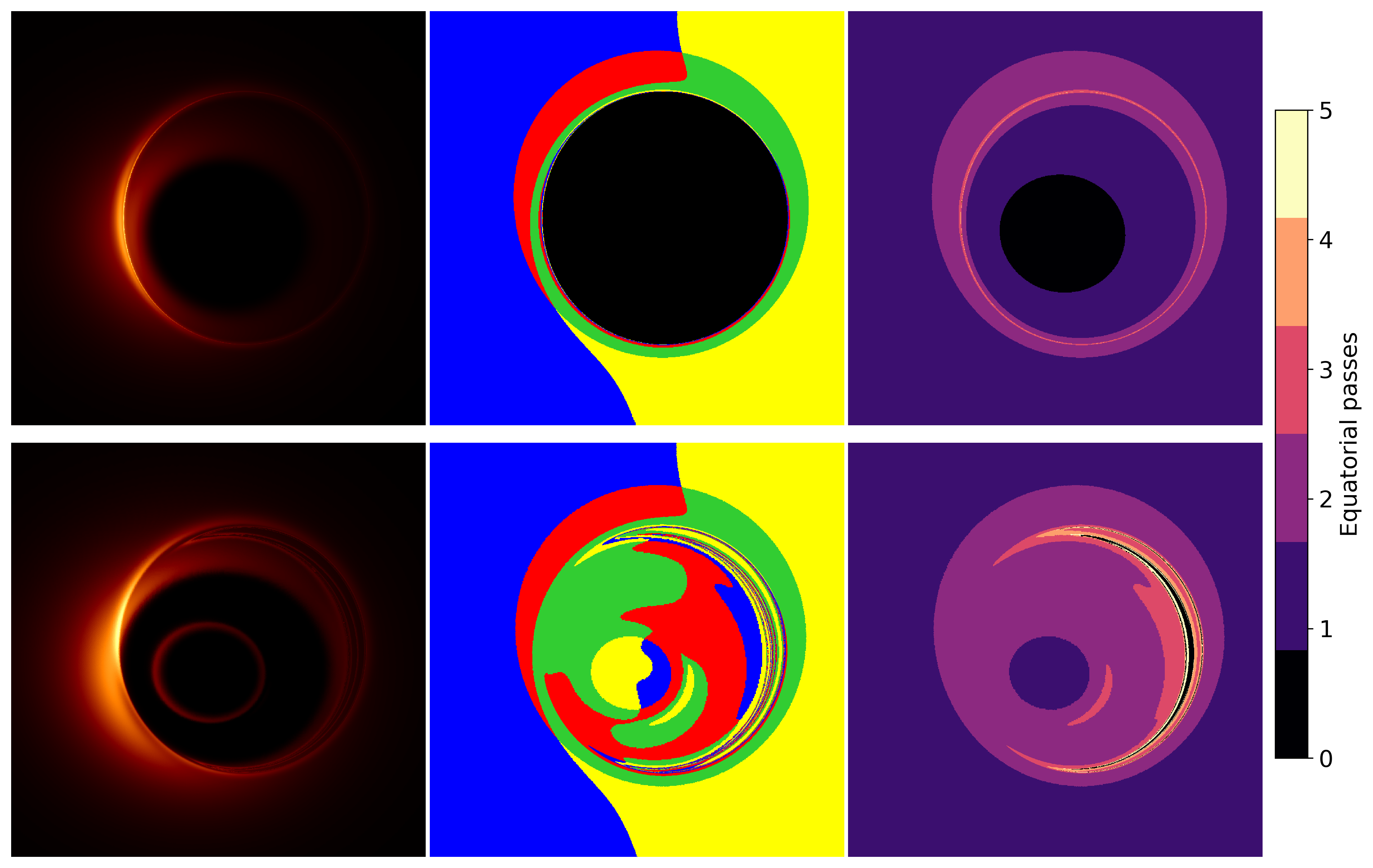}
    \caption{Comparison of a (\textit{top}) Kerr BH and (\textit{bottom}) a rotating BS (\texttt{C38}) at an inclination of 150\degree with respect to the rotation axis. From left to right, we show the observed intensity distribution, the four-colour screen and the number of equatorial passes. The colour scale for the number of equatorial passes has been capped at 5 for the BS, to facilitate comparison: the black arc in this image correspond to geodesics that have a higher number of equatorial passes, rather than an event horizon (as is the case for the BH).}
    \label{fig: rot BS comparison}
\end{figure*}

In this section, we investigate to what extent the results found in the previous section extend to the case of the rotating BSs presented in Sec.~\ref{subsec: theory - boson stars} --- see Table~\ref{tab: BS models}. In particular, we assess whether the visibility function of a rotating BS can be well approximated by that of a Kerr BH with additional modulations due to mirror images of the BH photon rings.\\

Figure \ref{fig: rot BS comparison} compares the \texttt{C38} model with a Kerr BH of the same mass and angular momentum, as seen by an observer at an inclination of $\theta_o = 150^\circ$. The emission disc peaks at $\rho = 1.94\, M \approx 36.2\mu^{-1}$, the (co-rotating) ISCO radius for a Kerr BH with reduced angular momentum $J/M^2 = 0.95$. This value is sufficiently close to the BS MSCO extracted by \texttt{FOORT} (see Table~\ref{tab: BS models radii}). The fudge factor was set to 0.7 to mimic the effect of the optical thickness of the disc, with a maximum of 5 passes through the disc adding to the total intensity (see Eq.~\eqref{eq: I tot}). Just like in Sec.~\ref{subsec: results - spherical symmetry}, we fix $\sigma = \frac{M}{4}$ and $\gamma = -2$.\\

\begin{figure}
    \centering
    \includegraphics[width=0.99\linewidth]{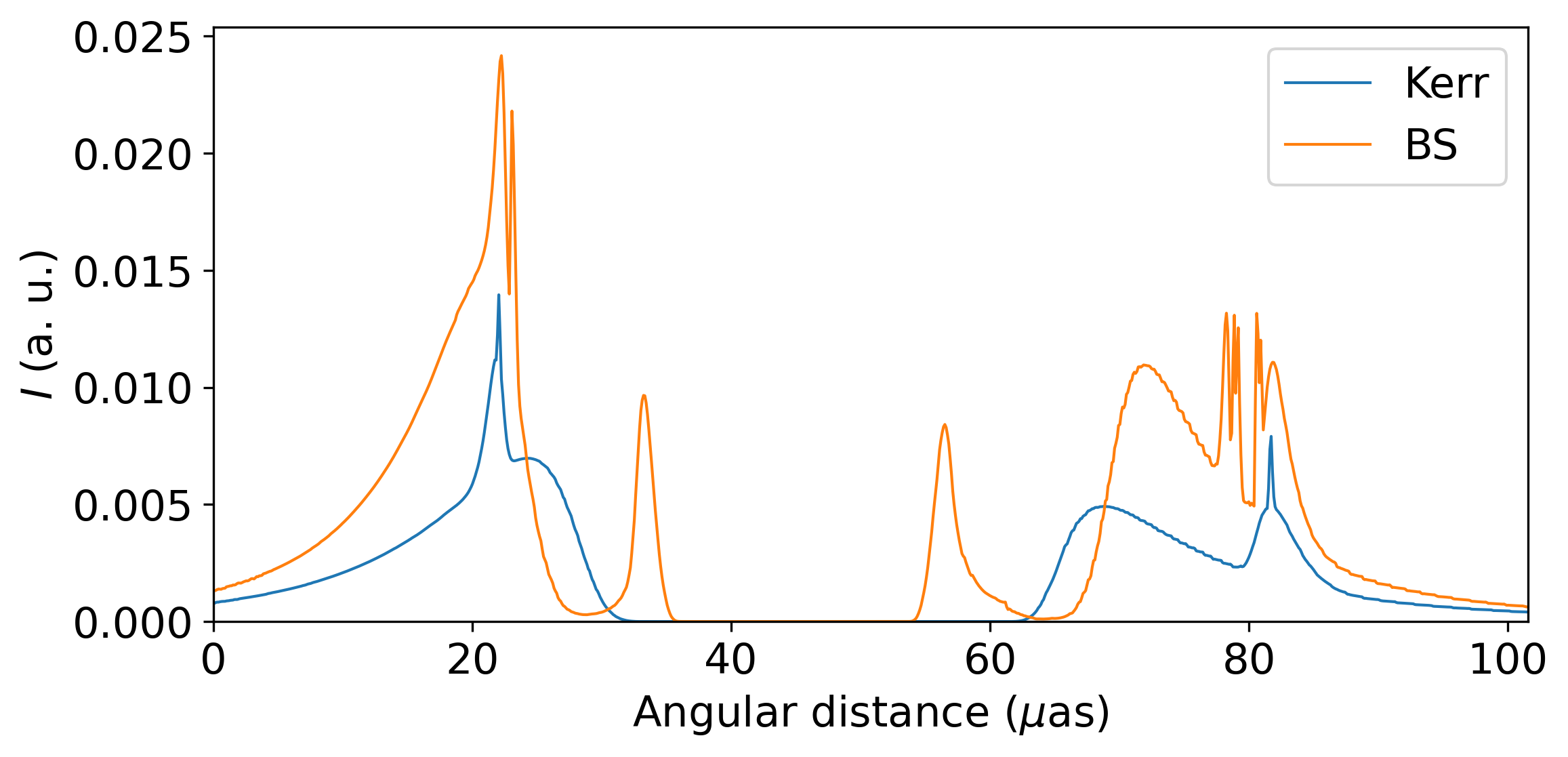}
    \caption{Intensity profile along a slice at $10^\circ$, for both the \texttt{C38} model and its Kerr counterpart in Fig.~\ref{fig: rot BS comparison}. The images are assumed to be $100\times100$ $\mu$as in size.}
    \label{fig: rot lineout comp C38}
\end{figure}

As expected, additional images appear in the intensity distribution of the BS compared to the BH. Again, a small central ring represents a lensed counterpart of the direct emission, not blocked given the absence of an event horizon. The first photon ring (contained in the $n=2$ equatorial passes region) appears to be comparable between the two images, but differences become manifest for higher order images. Figure~\ref{fig: rot lineout comp C38} shows a slice of the intensity map, along a straight line at an angle of 10$\degree$ with respect to the horizontal axis. The intensity profile for Kerr matches expectations, with the two photon rings appearing as exponentially smaller peaks on top of the broad direct emission. The corresponding profile for the BS shows a much more complex structure of superimposed images, as well as significant emission in the interior.

We note that the $n\geq 3$ region is crescent shaped, not being closed on the left-hand side of the image: this is not the case for Kerr spacetime. Indeed, as explained in Sec.~\ref{subsec: theory - light rings}, images of a Kerr BH (at any inclination) always contain a closed critical curve, parametrized by the radii at which unstable, spherical bound orbits exist (see e.g.~Fig.~2 in Ref.~\cite{Johnson:2019ljv}). In particular, the critical curve relevant to an observer at inclination $\theta_o$ is parametrized by $r\in [r^-_{\theta_o}, r^+_{\theta_o}]$, where $r^-_{\theta_o}$ ($r^+_{\theta_o}$) is the radius at which a prograde (retrograde) bound orbit exists with turning points in the $\theta$-direction at $\pm \theta_o$. In particular, the polar light ring that crosses the rotation axis always contributes to this critical curve.

We know that the rotating BS spacetimes considered here do not have corotating LRs, and the \texttt{C38} model appears to not have a polar light ring (see Supp.~Mat.~associated with Ref.~\cite{Siemonsen:2024snb}). Hence, if we naively extrapolate the above properties for Kerr, we would expect the `critical curve' for these rotating BSs to be `incomplete'; more precisely, we would expect the part of the critical curve that is sourced by corotating bound geodesics to be absent, i.e.~that part of the critical curve in the left-hand side of the image\footnote{That the emission on this side of the image is sourced by prograde null geodesics is apparent from the Doppler beaming.}. This would give rise to a crescent-shaped critical curve, and hence to crescent-shaped photon rings. Of course, this argument is not strictly valid: the geodesic equations for BSs are not separable in $r,\theta$, resulting in bound null geodesics whose radial coordinate varies along with the $\theta$-coordinate --- see e.g.~Ref.~\cite{Cunha:2016bjh}. Hence, we cannot simply parametrize the critical curve by the unique radius associated with a bound geodesic. However, we still hypothesize the interpretation for these \textit{incomplete photon rings} to be the following: they are incomplete due to the absence of suitable, unstable, bound, corotating null geodesics that are connected to infinity\footnote{In principle, these bound geodesics could live in \textit{pockets} (see e.g.~Ref.~\cite{Cunha:2016bjh}), but these would not source a sequence of photon rings, as they remain bound when perturbed.} in the rotating BS spacetimes. However, this statement would need to be confirmed by a more detailed analysis, e.g.~along the lines of Ref.~\cite{Cunha:2017eoe}, which additionally could help interpret the tips of the crescent-shaped regions.

Finally, we note that, while the overall lensing structure of a rotating BS spacetime can indeed be qualified as chaotic (see Ref.~\cite{Cunha:2016bjh} and the four-colour map in Fig.~\ref{fig: rot BS comparison}), the intensity map obscures this to some extent.\\

\begin{figure}
    \centering
    \includegraphics[width=0.99\linewidth]{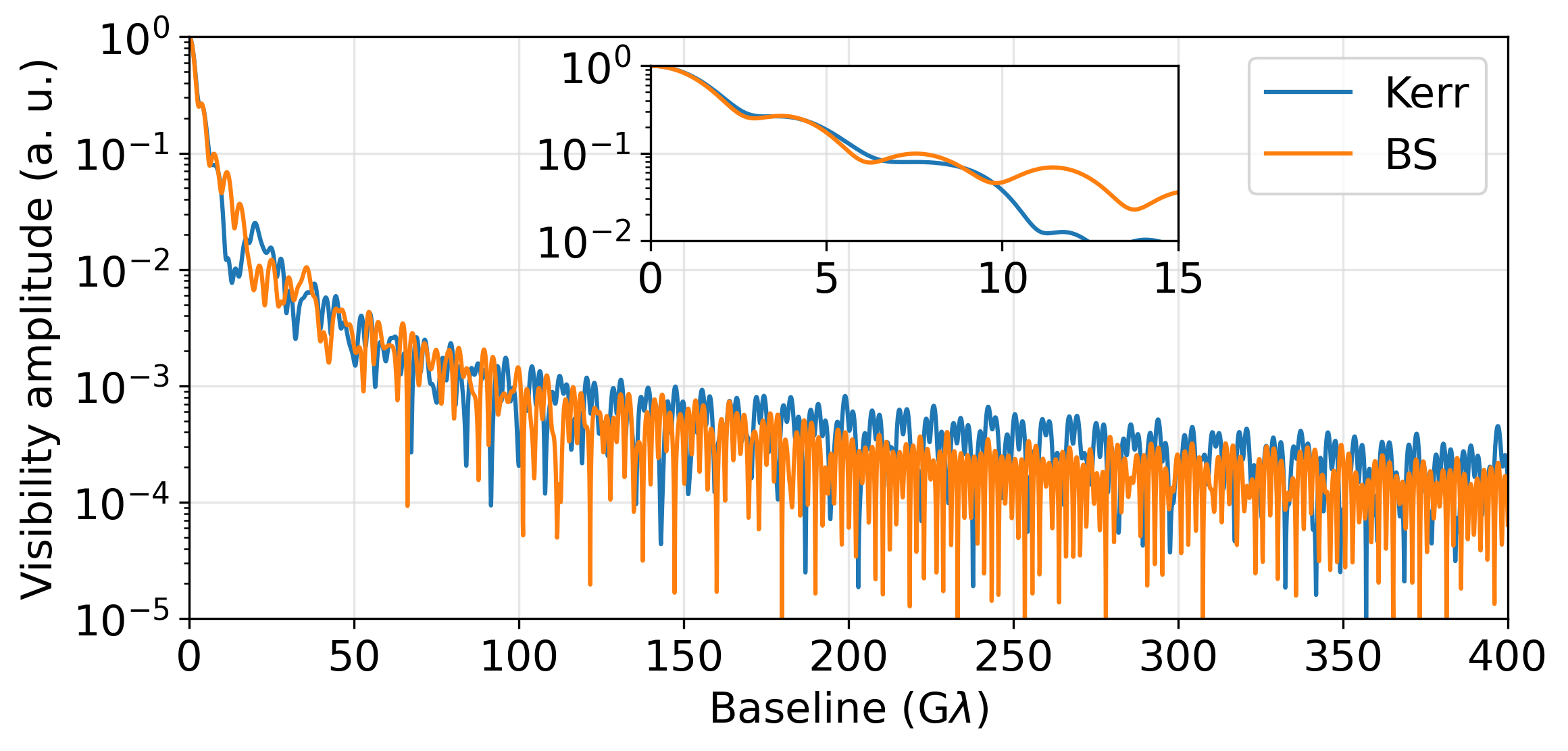}
    \caption{Visibility amplitude along a slice at $10^\circ$, for both the \texttt{C38} model and its Kerr counterpart. The baselines are scaled on the assumption that the images in Fig.~\ref{fig: rot BS comparison} are $100\times100$ $\mu$as in size.}
    \label{fig: rot bs visamp comp}
\end{figure}

Turning to the visibility amplitudes, it is now not surprising that the picture is more complicated than in the spherically symmetric case. Figure \ref{fig: rot bs visamp comp} shows the extracted visibility amplitudes for the rotating BS and its Kerr counterpart, along a $10^\circ$-slice in the Fourier plane. In this case, no additional modulation over long baselines is apparent in the case of the rotating BS, when compared to Kerr: this is not surprising, as we don't have clean pairs of closed photon rings in the image of the BS. The inset shows that the first (and to some extent, the second) dip in the VAs is comparable, however, which can be understood as the large-scale structure of both images being very similar. Having set the masses and angular momenta to be identical, this means that the overall size of the emitting region is comparable. More significant differences appear at larger baselines.

As noted in the Supp.~Mat.~of Ref.~\cite{Siemonsen:2024snb}, the \texttt{C39} and \texttt{C41} models have (at least) 1 and 2 polar LRs, respectively. Hence, we would expect an observer close to the rotation axis to see multiple sequences of photon rings in these cases. Additionally, the image for such a polar observer inherits the axisymmetry from the spacetime, such that the photon rings must be circular and closed. 

\begin{figure}
    \centering
    \includegraphics[width=\linewidth]{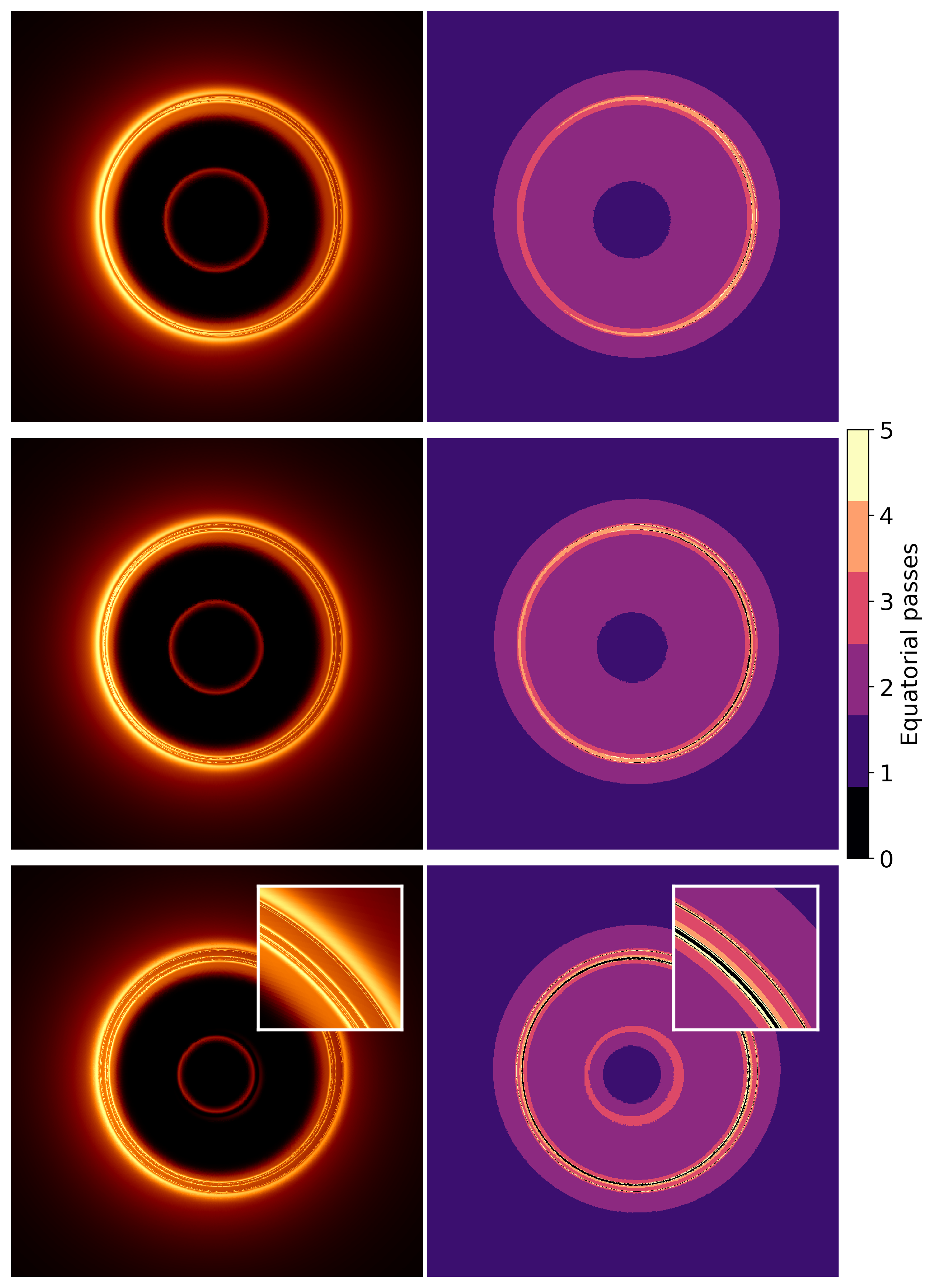}
    \caption{From top to bottom: the \texttt{C38}, \texttt{C39} and \texttt{C41} models, as seen by a near-polar observer at an inclination of $5^\circ$. The left column shows the intensity map, and the right column the number of equatorial passes $n$. Regions with $n>5$ are shown in black, to keep the color scale consistent. Insets are shown for \texttt{C41}, to visualize the ring structure in more detail.}
    \label{fig: rot bs on axis}
\end{figure}

This is shown in Fig.~\ref{fig: rot bs on axis}, where the rotating BSs are viewed at a $5^\circ$ inclination\footnote{An observer that is exactly on axis would cause numerical issues in the image, associated with the polar singularity in $\theta$.}  where indeed many circular photon rings are seen. In the case of the \texttt{C38} model, which does not have a polar LR, the $n\geq 4$ region is not a closed ring. However, the lensing is still sufficiently strong for circular photon rings to appear.
The photon rings of the \texttt{C41} model show a complex structure, with seemingly two critical curves being approached (the black bands, i.e.~$n>5$ regions, in Fig.~\ref{fig: rot bs on axis}), in line with the observation that this model appears to have two polar LRs. The insets for this model zoom in on the rings, with 4 times higher resolution.

\begin{figure}
    \centering
    \includegraphics[width=0.8\linewidth]{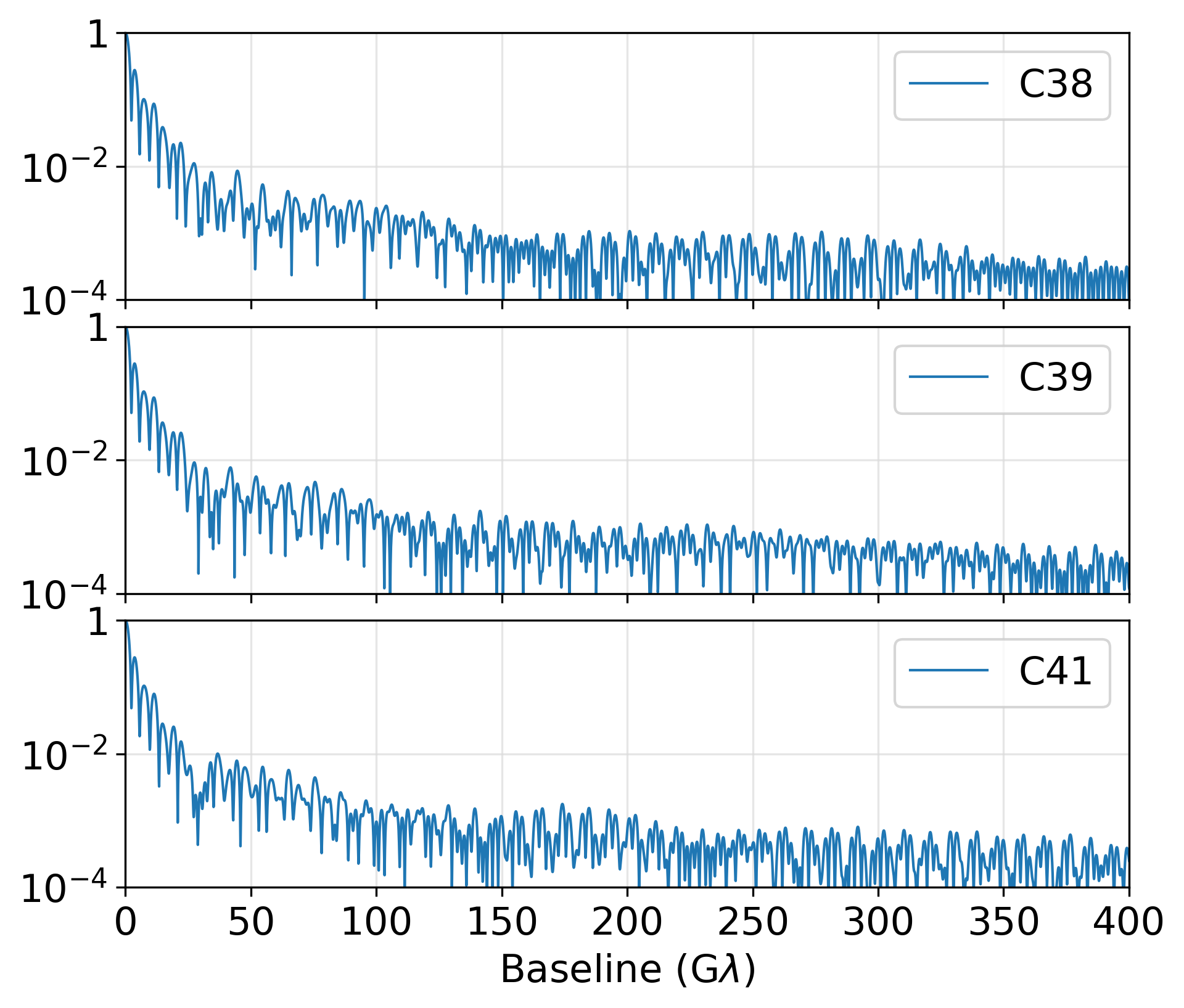}
    \caption{Visibility amplitudes corresponding to the images in Fig.~\ref{fig: rot bs on axis}, along a slice at an angle of 10$^\circ$.}
    \label{fig: rot bs on axis VA}
\end{figure}

The resulting VAs are shown in Fig.~\ref{fig: rot bs on axis VA}. We note that no strong low-frequency modulations are visible for any of the three models, despite a multitude of circular rings being present in the image. Although some low-frequency variation seems apparent, we can not link it to a specific pair of rings. A first possibility is that this is due to the fact that, again, many rings are present in the images, with no clear separation in scale and width as was the case in the spherically symmetric case. 

In the case of \texttt{C38} and \texttt{C39}, some crescent-shaped features are still present in the images\footnote{For \texttt{C38}, the $n=4$ region is not a closed ring. For \texttt{C39}, the ring is only broken for $n \gtrsim 5$: this likely means that, while a polar LR is present with $p_\phi = 0$, prograde bound geodesics with $p_\phi > \delta > 0$ (where $\delta$ is understood to be small) do not exist, or at least do not source nearly-bound geodesics.} --- this residual asymmetry is due to the observation angle of $5^\circ$, still slightly off-axis for numerical purposes.

For \texttt{C41}, Fig.~\ref{fig: rot bs on axis} clearly shows that the simple picture of exactly two rings of a certain width is not accurate here. The larger number of rings introduces more frequencies in the VA, which could be part of the reason that a clean modulation is absent.

We thus conclude that the absence of a co-rotating LR in the BS spacetime, and by extension a `complete photon shell', causes the photon ring structure to differ significantly from Kerr. The simple picture of a modulated visibility amplitude, as we have shown is the case in spherical symmetry under certain conditions, does not carry over to the rotating ultracompact BS studied in this work.

\subsection{Stalled disc model}
\label{subsec: results - stalled disc}

Notably, recent works \cite{Olivares:2018abq, Jaramillo:2026ygy} have shown that matter accreting onto a BS can (temporarily) be stalled at a radius $r_\Omega$ where the orbital frequency \eqref{eq: orbital angular frequency} for TCOs peaks. In this spirit, this section presents images assuming a disc with intensity peaking at $r_\Omega$. Additionally, App.~\ref{app: central disc model} also presents some results for a disc that extends all the way to the center of the BSs.\\

In the spherically symmetric case, Fig.~\ref{fig: Omega sph BS} tells us that the would-be peak of $\Omega$ for \texttt{A044} is located in between the two LRs, where no TCOs exist. In line with Refs.~\cite{Jaramillo:2026ygy, Olivares:2018abq}, however, we can assume that the stalling still takes place around the stable LR, as for smaller radii we have $\frac{{\rm d}\Omega}{{\rm d}r} > 0$. Our intensity prescription now has $\rho = r^s_{\rm LR} \approx 28\mu^{-1} \,, \gamma = -2$ and $\sigma = \frac{M}{2\mu}$: we choose the latter twice as large as our default value $\sigma = \frac{M}{4\mu}$ to retain significant support in the range where the MSCO-disc model resides. We extend the flow inwards from the MSCO, according to Cunningham's prescription\footnote{\label{foot: flow matching}This is not quite perfect: ideally, we would match this flow to the stable TCOs inside the stable LR, which is not done in the current version. However, this only applies to a small range of stable TCOs, as the emission decays inwards of the LR due to the decay constant $\gamma$ in Eq.~\eqref{eq: GLM profile}, and should hence give a decent approximation.} --- see App.~\ref{app: FOORT}.

\begin{figure}
    \centering
    \includegraphics[width=\linewidth]{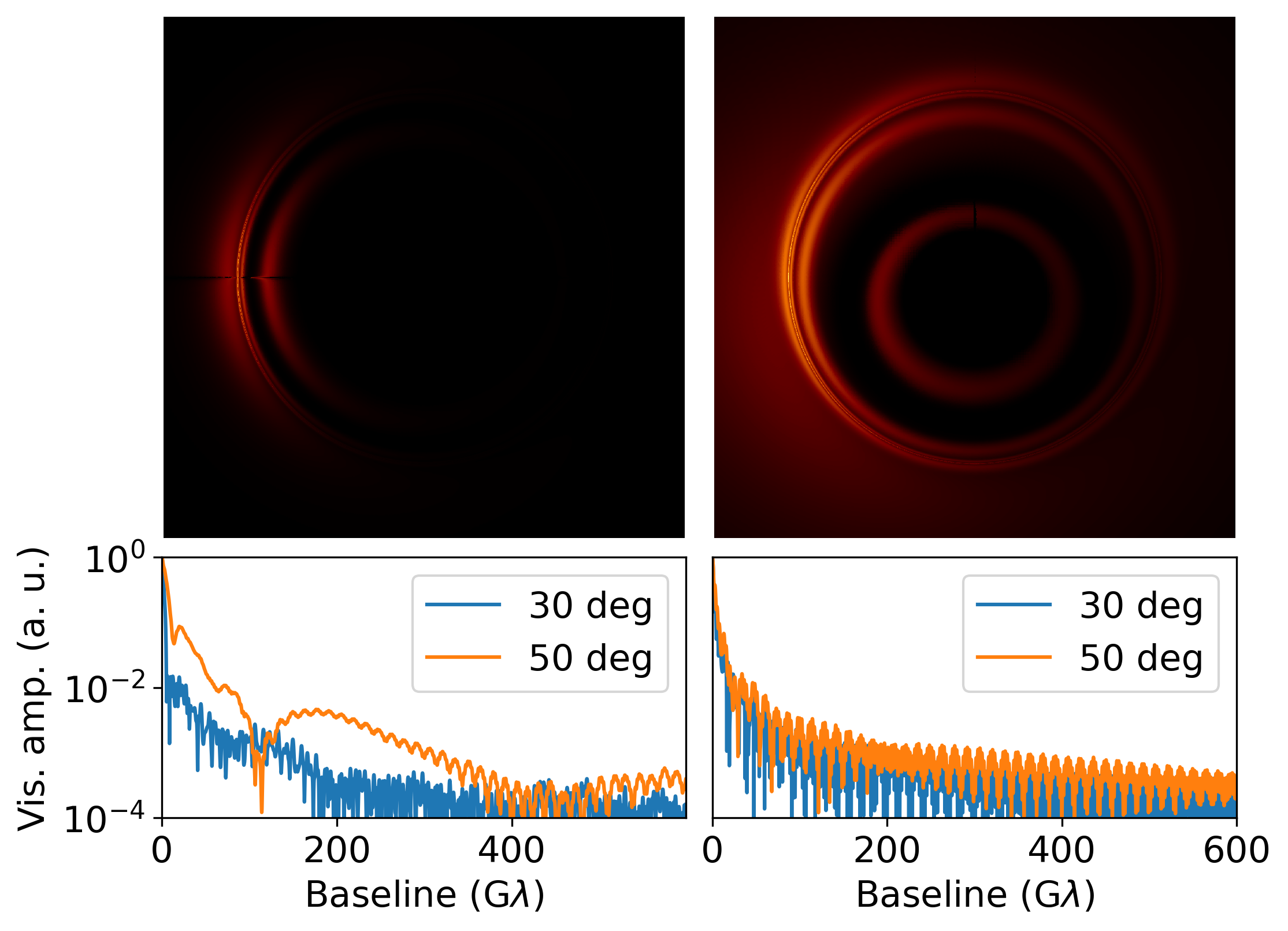}
    \caption{Ray-traced images of the \texttt{A044} BS surrounded by the stalled disc model, for (\textit{left}) an equatorial observer and a (\textit{right}) observer located at an inclination $\theta_0 = 150^\circ$ w.r.t. the vertical axis. The bottom panels show the total visibility amplitude as a function of baseline (assuming an image width of 100 $\mu$as) for slices along $30^\circ$ and $50^\circ$ in the visibility plane.}
    \label{fig: rOmega sph}
\end{figure}

Figure \ref{fig: rOmega sph} shows the images and some visibility amplitudes for the \texttt{A044} BS with the above disc model. The image for an equatorial observer is highly asymmetric due to strong Doppler beaming of the emitting matter. In the visibility, we see a clear bump at some angles, but no consistent modulation: this may be largely attributed to the high asymmetry, such that a uniform ring is no longer a good approximation for the photon rings. For the observer at $\theta_0 = 150\degree$, we see clear rings, but the first-order photon rings are less sharply defined. In other words, their separation is no longer large with respect to their width. There is no clear modulation in the total VA, nor in the VA of the first-order photon rings, as a result. The second-order rings are more sharply defined, but any signal and modulation we pick up from them is subdominant to the total signal.\\

For \texttt{C33}, we find $\mu r_\Omega^+ \approx 12.9$, such that $r_{\rm LR}^s < r_\Omega^+ < r_{\rm LR}^u$ --- see Fig.~\ref{fig: radial structure both}. We model the disc using $\rho = r_\Omega^+$ and $\sigma =  \frac{M}{2} \approx 3.1\mu^{-1}$. Again, we simply extend the flow inwards from the MSCO, and ignore the fact that for $r < r^+_\Omega$ the TCOs are stable --- see footnote \ref{foot: flow matching}. 
For \texttt{C38}, we find $\mu r_\Omega^+ = 27.8$. Again, this means that $r_{\rm LR}^s < r_\Omega^+ < r_{\rm LR}^u$. In this case, however, the MSCO is inside the unstable LR. Additionally, $r_\Omega^+ \approx r_{\rm MSCO}$, meaning that we don't expect significant differences with respect to the default setup, shown in Fig.~\ref{fig: rot BS comparison}, aside from the broader disc ($\sigma = \frac{M}{2\mu}$).

\begin{figure}[t]
    \centering
    \includegraphics[width=\linewidth]{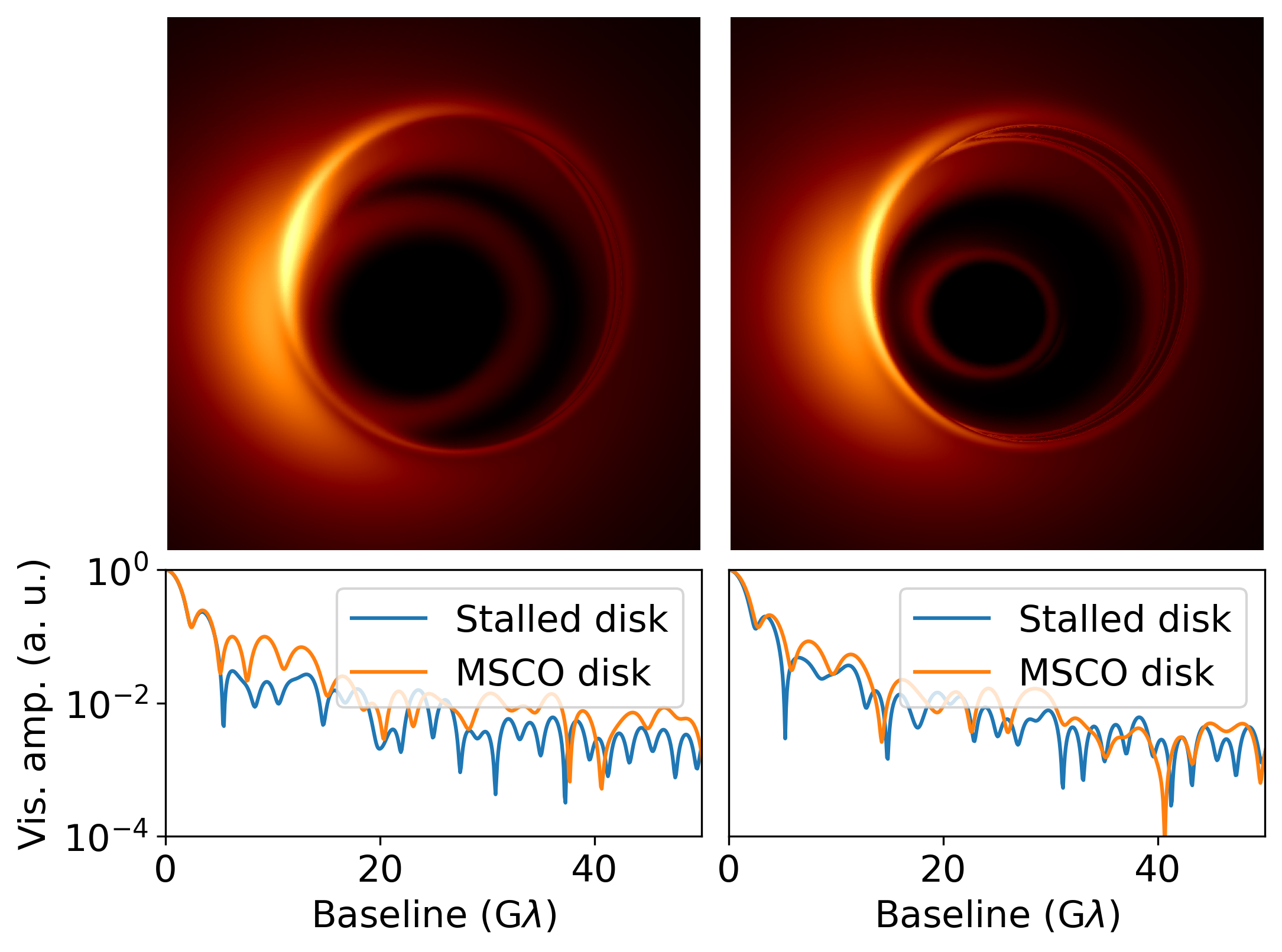}
    \caption{Ray-traced images of our (\textit{left}) \texttt{C33} and (\textit{right}) \texttt{C38} models, surrounded by stalled discs that peak around the radius $r^+_\Omega$, as described in the text. The observer is located at an inclination $\theta_o = 150^\circ$. The bottom panels compare the visibility amplitudes at baselines $\lesssim 50$ G$\lambda$ for these models and their MSCO-disc counterparts.}
    \label{fig: rotBS rOmega discs}
\end{figure}

We focus on the observer at $\theta_0 = 150\degree$, and show the resulting images in Fig.~\ref{fig: rotBS rOmega discs}. Compared to Figs.~\ref{fig: rot BS comparison} and \ref{fig: C33 comparison}, we see that the image has the same qualitative features, although the emission is more diffuse, which is reflected in the ring structures (especially the inner image of the \texttt{C33} model). This is largely due to the broader disc. We don't see strong differences in terms of the central brightness depression, and conclude that we don't expect fundamental differences in the VAs. The latter are plotted in the same Fig.~\ref{fig: rotBS rOmega discs} for baselines $\lesssim 50$ G$\lambda$.

\subsection{Lyapunov exponents}
\label{subsec: results - Lyapunov}

In this section, we briefly summarize the formulas that we obtained for the Lyapunov exponent \eqref{eq: Lyapunov} associated with equatorial light rings. The full derivation is given in App.~\ref{app: Lyapunov}, and we use these formulas to verify the \texttt{FOORT} implementation of the BS metrics in App.~\ref{app: light rings}.\\

In the BH imaging literature, the Lyapunov exponent $\gamma_L$ is defined with respect to the motion in the $\theta$-direction: an orbit is defined to be the trajectory from one turning point in $\theta$ back to itself \cite{Johnson:2019ljv, Lupsasca:2024wkp, Deich:2023oox}: a half-orbit hence produces one passage through the equatorial plane --- and hence, typically, the accretion disk. For spherically symmetric spacetimes, one orbit as defined above corresponds to a change $\Delta \phi=2\pi$ in the azimuthal angle, due to the motion being planar. Hence, defining an orbit as one period in $\phi$ is equivalent: this is no longer true for general axisymmetric spacetimes.

In spherical symmetry, the Lyapunov exponent for the Schwarzschild BH is (in our conventions) $\gamma_L = \pi$. For a general spherically symmetric spacetime \eqref{eq: sph symm metric}, and in particular the spherically symmetric BSs, we find that
\begin{equation}
\label{eq: Lyapunov BS}
    \gamma_L = \pi\,  r^2 \sqrt{\frac{- W_{\rm LR}''}{2X^2}}\,\Bigg|_{r=r^u_{\rm LR}}\,,
\end{equation}
where
\begin{equation}
    W_{\rm LR}(r) =- \frac{\eta_{\rm LR}^2}{\alpha^2} + \frac{1}{r^2}\,,
\end{equation}
and $\eta_{\rm LR} = E_{\rm LR}/ L_{\rm LR}$ is the ratio of the conserved quantities associated with the unstable LR at $r^u_{\rm LR}$. As an example, for the \texttt{A044} model we find that $\gamma_L \approx 2.95$.

For Kerr, the Lyapunov exponent is a function of the radius $r$ within the photon shell. Evaluated at the radius of either unstable LR, we always have that $\gamma_L = \pi$ with respect to the motion in $\theta$ --- see e.g.~Refs.~\cite{Johnson:2019ljv, Lupsasca:2024wkp}. 
The rotating BSs in this work are implemented according to the metric \eqref{eq: rot metric}. They all have unstable, counterrotating LRs, for which the Lyapunov exponent is given by \cite{Deich:2023oox}
\begin{equation}\label{eq: Lyapunov rot BS}
    \gamma_L = \pi r\sqrt{-\frac{\partial_r^2 H_-}{\partial_\theta^2 H_-}}\Bigg|_{r=r^u_{\rm LR}}\,,
\end{equation}
where $H_-(r, \theta)$ is defined in Eq.~\eqref{eq: potentials axi}. As an example, for the \texttt{C38} model we find $\gamma_L \approx 3.139$, which is very close to the corresponding result for Kerr. This is in line with the Supp.~Mat.~of Ref.~\cite{Siemonsen:2024snb}, where the orbital frequency of the unstable LR is also found to be very close to the corresponding value for Kerr.

To verify the numerical implementation of the BS metrics, it is more convenient to calculate the Lyapunov exponent for the unstable LR with respect to the motion in $\phi$. As shown in App.~\ref{app: Lyapunov rot BS}, this value can be calculated as
\begin{equation}\label{eq: Lyapunov rot BS phi orbit}
    \gamma_L^\phi = \frac{\pi}{\eta_{\rm LR}}\sqrt{\frac{f r}{g\sqrt{l}}\,H_-''}\Bigg|_{r=r^u_{\rm LR}}\,,
\end{equation}
where primes denote radial derivatives.
As an example, for the \texttt{C38} model we find $\gamma_L^\phi \approx 4.08$.

Coming back to the link between QNM damping times and the Lyapunov exponent, upon which we briefly touched in Sec.~\ref{subsec: theory - photon rings}, a naive extrapolation of these results would suggest that QNMs for the \texttt{A044} BS could have a somewhat longer damping timescale than the corresponding ones in Schwarzschild spacetime. Those of the rotating models, however, could be very similar.
However, this picture is too simplistic, since 1) QNMs for BHs are defined with respect to an event horizon, which BSs do not have, and 2) the correspondence between QNM damping times and the Lyapunov exponent is only valid in the eikonal limit. QNMs for BSs have been studied in e.g.~\cite{Yoshida:1994xi, Macedo:2013jja}. The complexity of BS ringdown is illustrated in Ref.~\cite{Siemonsen:2024snb}.

\section{Conclusions}
\label{sec: conclusions}

In this work, we have investigated to what extent double photon rings (in BS images) leave a prominent imprint on the VAs through low-frequency modulations. For the spherically symmetric BSs considered in this work, we find that
\begin{enumerate}
    \item The visibility amplitudes for near-equatorial observers can display a notable low-frequency modulation on long baselines.
    \item This low-frequency modulation is present, despite the asymmetry in the image as a result of the fluid velocity.
    \item Observers close to the symmetry axis (perpendicular to the accretion plane) still see highly circular photon rings, but no strong modulations in the total VA are apparent. This could be due to the larger separation between the photon ring pairs, and the stronger decay of the signal towards longer baselines.
    \item Observers at intermediate angles are less likely to see modulations, as the photon rings are less circular and have varying separation. This appears to violate the simple picture used to derive the modulations.
    \item Despite not having LRs, some BSs just outside of the ultracompact regime can still source modulations as they are still sufficiently compact to generate a finite number of photon rings.
\end{enumerate}

We reiterate that we have focussed on the first-order photon ring(s), i.e.~those sourced by geodesics that pass through the accretion disc twice. The above conclusions may change when considering higher-order photon rings, as they are expected to satisfy the approximations in Sec.~\ref{subsec: theory - visibility amplitudes} better. However, they contribute even less to the total intensity, and may only be detectable at very long baselines, making them even more elusive.\\

We also present ray-traced images of ultracompact, rotating solitonic BSs to investigate whether such modulations could be present for rotating ECOs that mimick Kerr spacetime. We have focused on images as seen by an observer at $\theta_o = 150^\circ$. In the cases considered, we do not find strong signatures of such a modulation. This is due to the images having a more complex photon ring structure with respect to those of their spherically symmetric counterparts. In particular, at generic observer inclinations, the photon rings of the rotating BSs considered in this work are \textit{incomplete}: instead, crescent- or bow-shaped structures appear, as already seen in Ref.~\cite{Vincent:2015xta, Vincent:2020dij}. This significantly breaks the assumptions for Eq.~\eqref{eq: visamp ring approx 2}, meaning that the absence of the modulations is not surprising. Even when viewed by near-polar observers, which has the effect of circularizing all features in the images, the complexity of the lensing appears prohibitive in sourcing clean modulations over long baselines.

We have also generated images of the BSs assuming a \textit{stalled disc model}, where the intensity peaks near the radius $r_\Omega$ where the angular velocity has a local maximum. This is inspired by recent GRMHD simulations \cite{Jaramillo:2026ygy}, and we briefly discuss the differences with respect to the standard MSCO-disc model.
Finally, we have provided analytic expressions for the Lyapunov exponent associated with the unstable LR in the accretion plane, both for spherically symmetric and rotating BSs, in the convention typically used for VLBI imaging. We used these to verify the correct numerical implementation of the metrics in \texttt{FOORT}.\\

We reiterate that the images in this work are generated using simplified disc models, and realistic accretion will almost certainly introduce additional asymmetries that can prohibit clear modulations to persist in the total VA. However, it has also been shown that time-averaging can smoothen out local and temporal variations, which can again alleviate some of these issues. 

Additionally, we underline that the aim of this study is not to put forward these particular BSs as the ideal alternatives to BHs. Indeed, we repeat that --- aside from \texttt{C33} --- the rotating models used in this work are unstable outside of axisymmetry \cite{Siemonsen:2020hcg, Sanchis-Gual:2019ljs}. Instead, we use BSs to understand what properties photon rings around generic BH mimickers could have; in this case, how the additional photon rings could interefere with the outer ones in Fourier space. We do note that at least the \texttt{A044} and \texttt{C33} models have been evolved in fully non-linear simulations without showing signs of instability \cite{Marks:2025jpt, Evstafyeva:2025mvx}, making them stand out in the zoo of ECOs and BH mimickers.

\section{Acknowledgments}

S.J.S. is grateful to Nils Siemonsen and Will East for providing the rotating boson star metrics. S.J.S. acknowledges interesting discussions with Daniel Mayerson, Ulrich Sperhake, Gareth Marks, Tamara Evstafyeva, Avery Broderick and Bart Ripperda's group at CITA. The authors are grateful for the comments on a first draft of this paper, by U.S., N.S. and D.M. mentioned above. S.J.S. is supported by the Centre for Doctoral Training at the University of Cambridge funded through STFC (ST/W006812/1, 2882538).

\appendix

\section{Ray-tracing with FOORT}
\label{app: FOORT}

This section briefly summarizes the key features of our \texttt{Flexible Object-Oriented Ray Tracer} \cite{Mayerson:2025foo, MayersonFOORTgithub}.
The geodesic equations are integrated using the
Runge-Kutta 4 or the Verlet scheme to produce images of a variety of spacetimes. FOORT is highly flexible and easily accommodates new analytical or numerical spacetimes. It employs adaptive mesh refinement by refining the initially uniform grid in regions
where selected image features change significantly.

The BS metrics are implemented using interpolators that require independently constructed BS spacetimes. The spherically symmetric metric \eqref{eq: sph symm metric}
uses cubic splines following Ref.~\cite{kluge-splines}.
The rotating BS metrics are implemented using bicubic splines, whose implementation is accelerated with GitHub Copilot.
We also briefly comment on the resolution, and its effect on the estimates for the radii in Table \ref{tab: BS models radii}. The spherically symmetric BS metrics are obtained with high accuracy ($\Delta r \approx 10^{-2}$) in the shooting code. The rotating BS metrics are obtained using a compactified coordinate $x = r/(1+r)$: hence, the radial resolution is variable, and depends on the resolution $\Delta x$. For the \texttt{C38} model, we have $\Delta x = 1500^{-1}$, while for the others we have $\Delta x = 500^{-1}$. We can now deduce that e.g.~the (un)stable LR radius for \texttt{C38} in Table \ref{tab: BS models radii} is determined with an accuracy of about $\Delta r \approx  0.2$ ($\Delta r \approx 2$).\\

\texttt{FOORT} allows the user to specify a model for the velocity of the emitting matter in the disc, following e.g.~Refs.~\cite{Cardenas-Avendano:2022csp, Pu:2016qak}. At any point $p$, the velocities for a radially infalling geodesic and a circular geodesic (if it exists) are calculated. A mixture of the two can then be prescribed using the mixing parameters $\beta_r, \beta_\phi$, with optional sub-Keplerian rescaling (see
Eqs.~(B36),~(B50),~(B51) in Ref.~\cite{Cardenas-Avendano:2022csp}): note that changing these parameters will generally render the flow non-geodesic. In regions where circular geodesics do not exist, the flow is smoothly extended past the ISCO using Cunningham's prescription \cite{Cunningham:1975zz}, which assumes the conserved quantities of the ISCO and infers a radial velocity from the normalization condition. The final fluid velocity vector $w^\mu$, satisfying $w_\mu w^\mu = -1$, is then combined with the geodesic velocity $u^\mu$ to give the redshift factor 
\begin{equation}\label{eq: redshift factor}
    g(p) = \frac{-E}{w(p)_\mu u^\mu}\,
\end{equation}

\begin{figure}[t]
    \centering
    \includegraphics[width=1\linewidth]{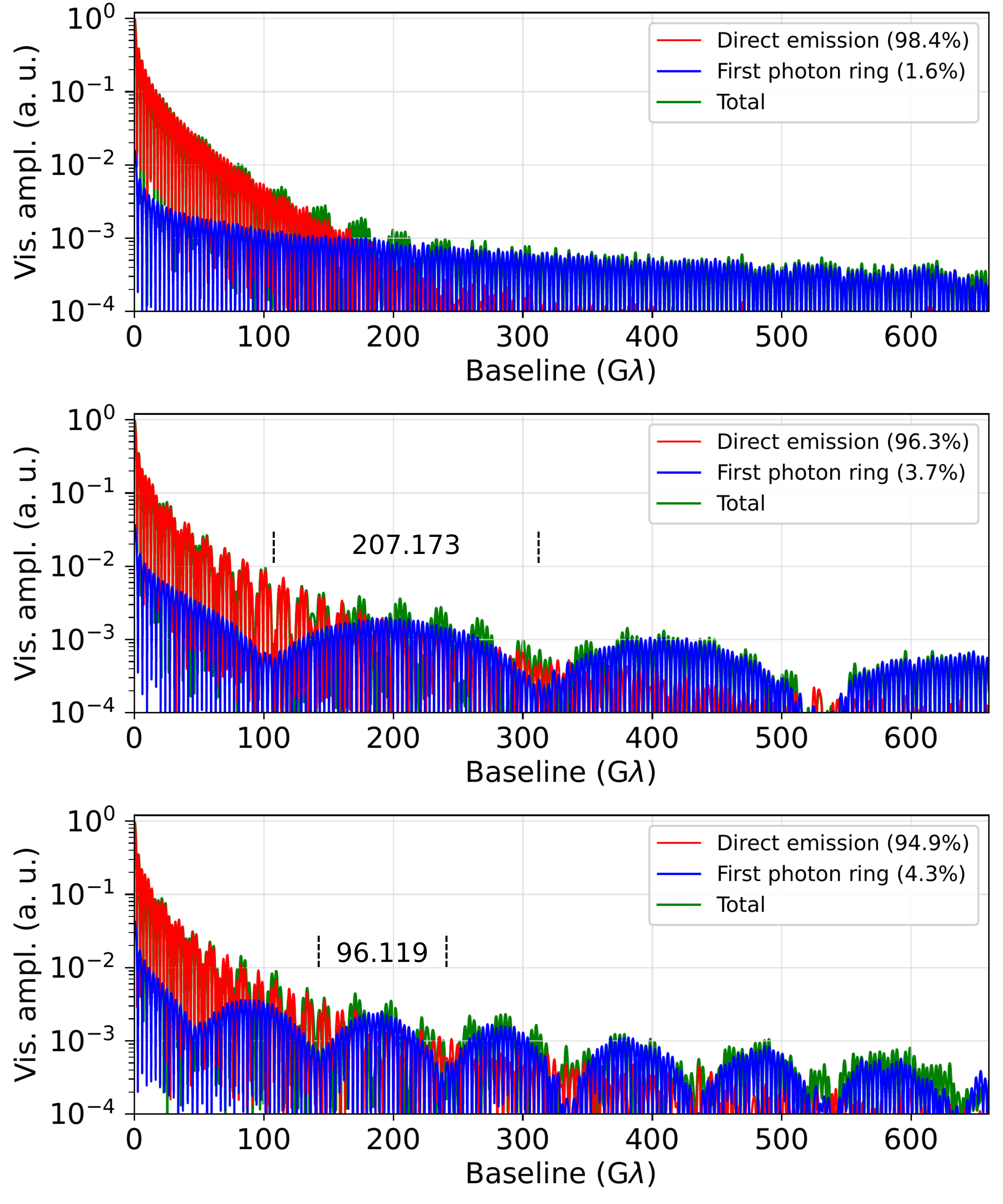}
    \caption{Same as Fig.~\ref{fig: visamps sph symm}, but assuming radially infalling matter in the disc rather than Keplerian.}
    \label{fig: radial fluid visamps}
\end{figure}

In this work, we assume a circular Keplerian flow, i.e.~$\xi = \beta_r = \beta_\phi = 1$. As an illustration, we show the VAs obtained for the \texttt{A044} model with a disc composed of radially infalling matter ($\beta_r = \beta_\phi = 0$) in Fig.~\ref{fig: radial fluid visamps}. While the high-frequency dips are now more pronounced due to the enhanced symmetry in the image, the overall picture is the same as in Fig.~\ref{fig: visamps sph symm}. In particular, the low-frequency modulations are still clearly visible.

\section{Lyapunov exponent}
\label{app: Lyapunov}

In this section, we derive Eq.~\eqref{eq: Lyapunov BS} for a generic spherically symmetric spacetime \eqref{eq: sph symm metric} as well as Eqs.~\eqref{eq: Lyapunov rot BS} and ~\eqref{eq: Lyapunov rot BS phi orbit} for the rotating, axisymmetric spacetime \eqref{eq: rot metric}. For simplicity, we will drop the subscript LR on quantities associated with the unstable LR, e.g.~$r^u \equiv r^u_{\rm LR}$.

\subsection{Spherical symmetry}
\label{app: Lyapunov sph symm}

If the motion is confined to the plane $\theta = \frac\pi2$, we cannot count passages through this plane to determine the Lyapunov exponent $\gamma_L$. Due to spherical symmetry, however, we can equivalently calculate $\gamma_L$ with respect to the $\phi$-motion, and count the number of times the azimuthal angle $\phi$ has increased by $\pi$.

Consider a nearly-bound null geodesic in the equatorial plane, starting close to the unstable LR at radial coordinate $r_0 = r^u + \delta r_0$ with $\delta r_0 \ll r_{\rm LR}^u$ (and WLOG we can assume $\phi_0 = 0$). In this context, nearly bound means that the conserved quantities $E = -p_t, L = p_\phi$ of the geodesic are approximately those associated with the LR, which we denote as $E^u, L^u$. From the normalization condition $p^\mu p_\mu = 0$, we derive that the geodesic is subject to the equation
\begin{equation}
   X^2 \left(\frac{{\rm d}r}{{\rm d}\lambda}\right)^2 - \frac{E^2}{\alpha^2} + \frac{L^2}{r^2} = 0\,,
\end{equation}
where $\lambda$ is an affine parameter.
We can rewrite this as
\begin{equation}
    \left(\frac{{\rm d}r}{{\rm d}\lambda}\right)^2 = \frac{-W(r)}{X^2} L^2\,,
\end{equation}
where we defined the potential
\begin{equation}
    W(r) = -\frac{E^2}{L^2\, \alpha^2} + \frac{1}{r^2}\,.
\end{equation}
As we will focus on nearly-bound geodesics that diverge from the LR towards infinity, we can take the positive root and write
\begin{equation}
    \frac{{\rm d}r}{{\rm d}\lambda} \sqrt{\frac{X^2}{-W(r)L^2}} = 1\,.
\end{equation}
Using the $\phi$-equation,
\begin{equation}
    \frac{{\rm d}\phi}{{\rm d}\lambda} = \frac{L}{r^2}\,,
\end{equation}
we now find that
\begin{equation}
    \frac{{\rm d}r}{{\rm d}\lambda} \sqrt{\frac{X^2}{-W(r)}} = {\rm sign}(L)\, \frac{{\rm d}\phi}{{\rm d}\lambda} r^2\,.
\end{equation}
We can now integrate this expression along the geodesic from $\lambda_0$ to $\lambda_n$, where the latter is chosen such that $\phi$ has increased by $n$ half-orbits, i.e.~$\phi_n = n\pi$. This necessarily means that $L>0$, giving that
\begin{equation}
    \int_{\lambda_0}^{\lambda_n}  \sqrt{\frac{X^2}{-W(r)}} \frac{{\rm d}r}{{\rm d}\lambda}\, {\rm d}\lambda = \int_{\lambda_0}^{\lambda_n} r^2\, \frac{{\rm d}\phi}{{\rm d}\lambda} \, {\rm d}\lambda\,.
\end{equation}
Denote the radius at $\lambda_n$ as $r^u + \delta r_n$, and assume $\delta r_n \ll r^u$ still holds. We can then approximate the right-hand side as
\begin{equation}
    \int_{r_0}^{r_n}  \sqrt{\frac{X^2}{-W(r)}} {\rm d}r \approx (r^u)^2\, n \pi.
\end{equation}
Recall that for the LR and its conserved quantities $W^u(r^u) = (W^u)'(r^u) = 0$, such that we can approximate and Taylor expand
\begin{align*}
    W(r) & \approx W^u(r) \\
    & \approx \frac12 (W^u)'' (r^u) (r-r^u)^2\,.
\end{align*}
This allows us to approximate the left-hand side as
\begin{align*}
    \int_{r_0}^{r_n}  \sqrt{\frac{X^2}{-W(r)}} {\rm d}r & \approx \int_{r_0}^{r_n}  \sqrt{\frac{2X^2(r^u)}{-(W^u)'' (r^u) (r-r^u)^2}} {\rm d}r \\
     & \approx \sqrt{\frac{2X^2(r^u)}{-(W^u)'' (r^u)}} \ln\left(\frac{\delta r_n}{\delta r_0}\right)\,.
\end{align*}
From Eq.~\eqref{eq: Lyapunov}, we have that
\begin{equation}
    \gamma_L = \frac{1}{n} \ln\left(\frac{\delta r_n}{\delta r_0}\right)\,,
\end{equation}
from which we now find Eq.~\eqref{eq: Lyapunov BS}.

\subsection{Rotating BS}\label{app: Lyapunov rot BS}

As the LR is confined to the equatorial plane, the Lyapunov exponent with respect to the motion in $\theta$ cannot be directly extracted. Ref.~\cite{Deich:2023oox} provides a formula that circumvents this, and acts as the limit for non-equatorial bound geodesics. We apply this expression to the metric \eqref{eq: rot metric}, and hence find the Lyapunov exponent for the unstable LR in the rotating BS spacetime. To test the metric implementation in \texttt{FOORT}, however, we can calculate the Lyapunov exponent with respect to the motion in $\phi$, and compare that to a numerically evolved geodesic.

\subsubsection{With respect to the $\theta$-motion}

Ref.~\cite{Deich:2023oox} shows that the Lyapunov exponent for the unstable LR with respect to the $\theta$-motion is given by
\begin{equation}\label{eq: Lyap deich}
    \lambda = \pi \sqrt{-\frac{g_{\theta\theta}}{g_{rr}}\frac{\partial_r^2 V_{\rm eff}}{\partial_\theta^2 V_{\rm eff}}}\Bigg|_{r = r^u}\,,
\end{equation}
where $V_{\rm eff}(r,\theta)$ for null geodesics is defined by
\begin{equation}
    0 = V_{\rm eff} + \frac{p_r^2}{g_{rr}}\,.
\end{equation}
We immediately have that $\frac{g_{\theta\theta}}{g_{rr}} = r^2$. From the normalization condition $p_\mu p^\mu = 0$ for the metric \eqref{eq: rot metric}, we find that equatorial null geodesics, parametrized by $\lambda$, obey
\begin{equation}\label{eq: radial motion axi}
    \left(\frac{{\rm d}r}{{\rm d}\lambda}\right)^2 =\frac{L^2}{gl} (\eta - H_+)(\eta - H_-)\,,
\end{equation}
where the potentials are given in Eq.~\eqref{eq: potentials axi},
and $\eta = E / L$ is the ratio of conserved quantities.
Hence, we have that
\begin{equation}\label{eq: Veff motion axi}
    V_{\rm eff} =\frac{L^2}{f} (\eta - H_+)(\eta - H_-)\,.
\end{equation}
The retrograde, unstable LR satisfies
\begin{equation}\label{eq: unst retro LR}
    H_-(r^u) = \eta\,, \partial_r H_-(r^u) = 0\,,\partial_r^2H_-''(r^u) > 0
\end{equation}
in the equatorial plane.
Hence, the expression \eqref{eq: Lyap deich} simplifies, as the second derivatives need to act on the factor $(\eta - H_-)$ to give non-zero contributions. Hence, the final expression is indeed Eq.~\eqref{eq: Lyapunov rot BS}, as the remaining prefactors cancel out.

\subsubsection{With respect to the $\phi$-motion}

To obtain the Lyapunov exponent with respect to the $\phi$-motion, we instead proceed like we did in App.~\ref{app: Lyapunov sph symm}. We raise the index on $p_\phi = L$ to find that the motion in the $\phi$-direction is governed by
\begin{equation}
    \label{eq: phi motion rot BS}
    \frac{{\rm d}\phi}{{\rm d}\lambda} =-\frac{L}{f} \left(H_+H_- - \frac\eta2(H_+ + H_-)\right)\,.
\end{equation}
While the nearly-bound geodesic is still sufficiently close to the unstable LR, we can use Eq.~\eqref{eq: unst retro LR} to approximate this as
\begin{equation}
    \label{eq: approx phi motion rot BS}
    \frac{{\rm d}\phi}{{\rm d}\lambda} \approx \frac{L\eta}{2f} \left(\eta - H_+\right)\,.
\end{equation}
As we are looking at retrograde motion, i.e.~$L<0$, we can rewrite Eq.~\eqref{eq: radial motion axi} as
\begin{equation}
    L = - \frac{{\rm d}r}{{\rm d}\lambda} \sqrt{\frac{gl}{(\eta - H_+)(\eta - H_-)}}\,,
\end{equation}
which can be combined with Eq.~\eqref{eq: approx phi motion rot BS} to find
\begin{equation}
    \frac{{\rm d}r}{{\rm d}\lambda} (H_- -\eta)^{-1/2} = -\frac{2f}{\eta} \left[-gl(\eta-H_+)\right]^{-1/2} \frac{{\rm d}\phi}{{\rm d}\lambda}\,.
\end{equation}
Note that $\eta - H_+ = \frac{-2f}{r\sqrt{l}}$ at the unstable LR. Integrating these equations from $\lambda_0$ to $\lambda_n$, i.e.~from $r_0$ to $r_n$ such that $\phi$ has decreased from 0 to $-n\pi$ (as the motion is retrograde), we find that
\begin{equation}
    \int_{r_0}^{r_n}\frac{{\rm d}r}{\sqrt{(H_- -\eta)}} \approx n\pi \frac{\sqrt2}{\eta}\sqrt{\frac{fr^u}{g\sqrt{l}}}\Bigg|_{r=r^u}\,.
\end{equation}
We now can approximate the integral by noting that, while the divergence is still small ($\delta r_n \ll r^u$), Eq.~\eqref{eq: unst retro LR} implies
\begin{equation}
    H_- \approx \eta + \frac12H_-''(r^u) (r-r^u)^2\,,
\end{equation}
where primes now denote radial derivatives, such that
\begin{equation}
    \frac1n\ln\left(\frac{\delta r_n}{\delta r_0}\right)\sqrt{\frac{2}{H_-''(r^u)}} \approx \pi \frac{\sqrt2}{\eta}\sqrt{\frac{fr^u}{g\sqrt{l}}}\Bigg|_{r=r^u}\,.
\end{equation}
Rewriting now gives Eq.~\eqref{eq: Lyapunov rot BS phi orbit}.

Note that, in case of the Kerr metric, Lyapunov exponents can be calculated for any bound null geodesic that makes up the photon shell --- see e.g.~Refs.~\cite{Lupsasca:2024wkp, Kapec:2019hro}. Whether a similar calculation is possible remains to be seen, as the non-separability of the geodesic equations (and the lack of analytic solutions) make the analogous treatment impossible. A more thorough investigation of this problem is left for future endeavours.

\section{Numerical validation: Light rings}
\label{app: light rings}

In this section, we aim to provide some verifications of the correct implementation of the spacetime  (described in App.~\ref{app: FOORT}).
To do so, we numerically evolve geodesics that approximate the light rings present in the spacetimes of interest. In the case of stable light rings, we should find that the radial coordinate of the geodesic oscillates around the initial value, despite discretisation errors and numerical integration. For the unstable light rings, we expect the radial coordinate to diverge from its inital value, as discretisation implies that we are unable to initialise the geodesic at the unstable light ring itself. The divergence should then (at least initially) be governed by an exponentially growing drift, at a rate set by the Lyapunov exponent --- see Sec.~\ref{subsec: theory - photon rings} and App.~\ref{app: Lyapunov}.

\begin{figure}
    \centering
    \includegraphics[width=\linewidth]{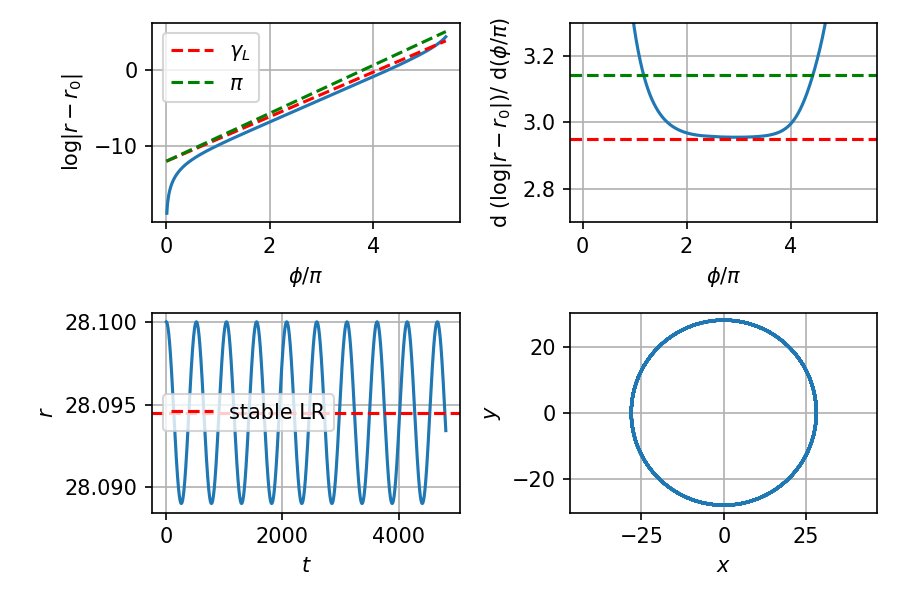}
    \caption{(\textit{top}) Radial growth of a geodesics initialized close ($\delta r_0 = 10^{-7}$) to the unstable LR, as a function of the angle $\phi$. The panel on the right shows the rate of exponential growth, which is seen to agree with the predicted Lyapunov exponent over about a full orbit, in the range $2\pi \lesssim \phi \lesssim 4\pi$. (\textit{bottom}) Radial evolution of a geodesic initialized close to the stable LR. The panel on the right shows the orbit in the equatorial plane.}
    \label{fig: test BS LRs}
\end{figure}

To test the implementation of the spherically symmetric BS metric, we use the representative \texttt{A044} model. The Lyapunov exponent \eqref{eq: Lyapunov BS} for the unstable LR is approximately 2.95. Fig.~\ref{fig: test BS LRs} shows the evolution of two null geodesics, initialized close to the LRs. The top row shows that, a geodesic initialized at $r_0 = r_{LR}^u + 10^{-7}$ experiences a regime in which the deviation is exponential, at a rate predicted by the Lyapunov exponent, as expected. The stronger initial deviations are likely influenced by discretization, whereas the strong deviation at late times is due to the geodesic leaving the regime of validity for Eq.~\eqref{eq: Lyapunov}.
On the other hand, the geodesic initialized close to the stable LR experiences a small radial oscillation, as expected.

\begin{figure}
    \centering
    \includegraphics[width=\linewidth]{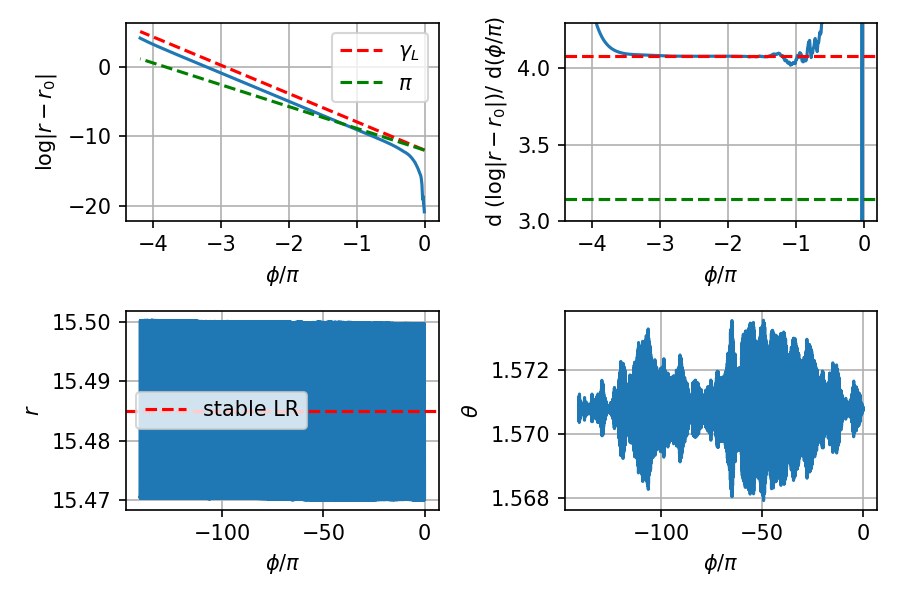}
    \caption{(\textit{top}) Radial growth of a geodesics initialized close ($\delta r_0 = 10^{-6}$) to the unstable LR, as a function of the angle $\phi$ (note that the geodesic is counterrotating, hence $\phi$ is decreasing). The panel on the right shows the rate of exponential growth, which is seen to agree with the predicted Lyapunov exponent $\gamma_L^\phi$ over about a full orbit, in the range $-7\pi/2 \lesssim \phi \lesssim -3\pi/2$. (\textit{bottom}) Radial evolution of a geodesic initialized close to the stable LR (rapidly oscillating). The panel on the right shows the $\theta$-coordinate along the geodesic, showing that some numerical jitter is present.}
    \label{fig: test rot BS LRs}
\end{figure}

We perform a similar test for the \texttt{C38} rotating BS, and show the results in Fig.~\ref{fig: test rot BS LRs}.
From the numerical data, we find that the stable LR is located at $r \approx 15.5$, with local resolution $\Delta r \lesssim 0.2$. The unstable LR is located around $r_u\approx 56$ at a resolution of $\Delta r \approx 2$. Integrating (intially circular) geodesics close to these radii, we are able to recover the expected behavior in \texttt{FOORT}.
Numerically, we find that the exponential growth of the radius associated with the diverging geodesic close to the unstable LR (numerically finetuned to start at $r = 56.507306$) is in excellent agreement with the semi-analytically obtained $\gamma_L^\phi \approx 4.08$. To get clean behavior close to the stable LR, we had to decrease the adaptive step size for the integrator from the default 0.03 to 0.01: not doing so would eventually cause a small drift away from the stable LR\footnote{We use the default value of 0.03 for all experiments in the main text. We have verified that changing this value to 0.01 did not have any notable effect on our conclusions and images there.}. Experiments then suggest that the stable LR is located around $r_s\approx 15.485$ (in the interpolated metric, at least). Fig.~\ref{fig: test rot BS LRs} also shows that, aside from some jittering, the geodesic remains close to the equatorial plane. We also note that throughout the integration, the invertibility of the metric is monitored to be satisfied to at least $10^{-13}$.

\section{Convergence}\label{app: convergence}

This section provides a brief analysis to show that the VAs presented in the main text are robust and informative up to $\gtrsim 10^{-4}$.

\begin{figure}
    \centering
    \includegraphics[width=
    \linewidth]{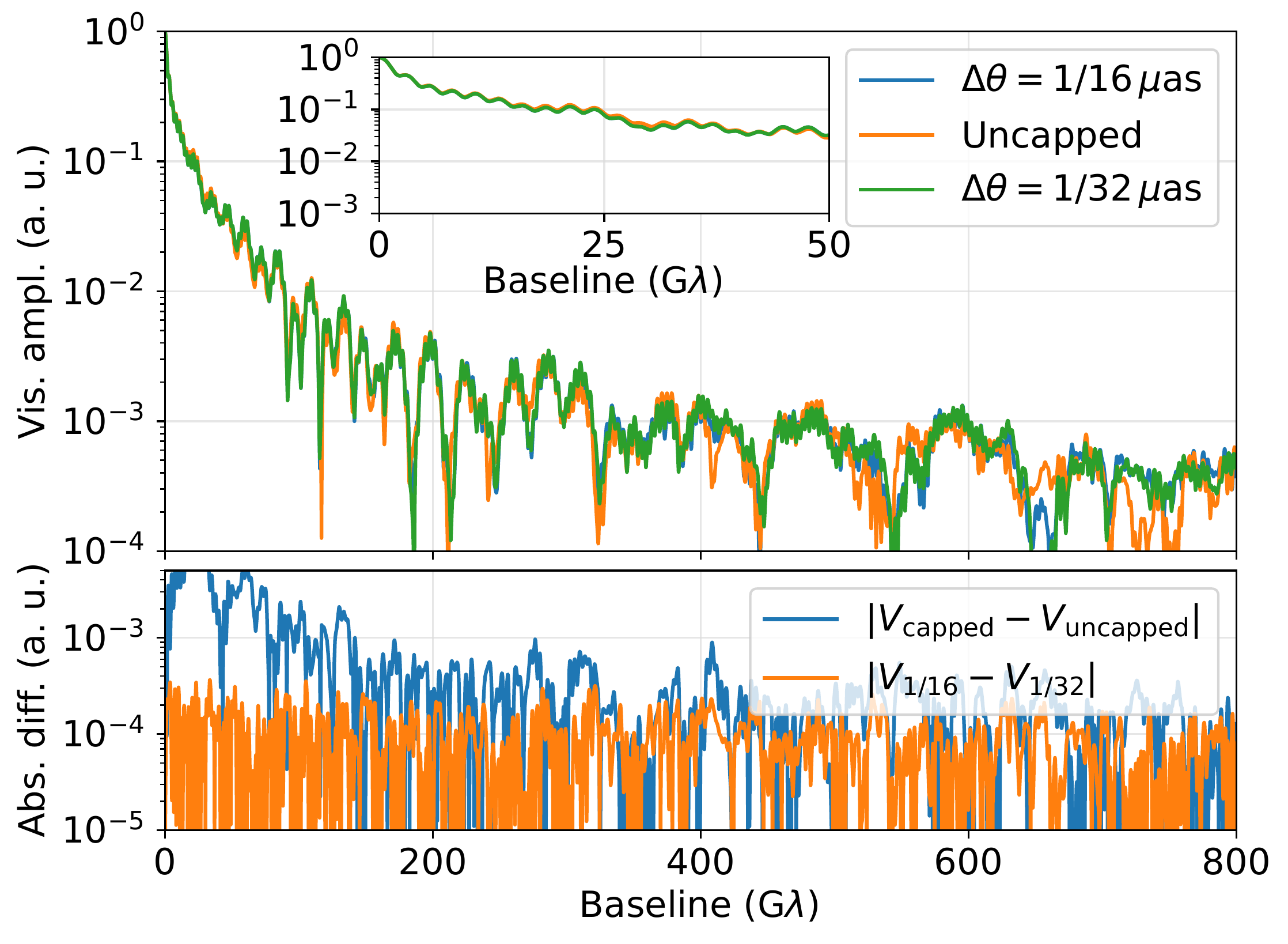}
    \caption{Comparison of the normalised visibility amplitudes obtained at different resolutions, for the \texttt{A044} BS with the MSCO disc, for an equatorial observer. The image size is $100\times100$ $\mu$as, and $\Delta \theta$ is the angular separation between neighbouring pixels. (\textit{top}) We plot the visibility amplitude for 1) a low-resolution simulation, capped at $1\times10^6$ geodesics, 2) the same simulation without a cap on the number of geodesics and 3) a simulation with doubled resolution, capped at $4\times10^6$ geodesics. (\textit{bottom}) We plot the absolute differences between the simulations at different resolution, i.e.~adding a sublevel, and the difference between the low-resolution sim}
    \label{fig: resolution BS}
\end{figure}

We start with the spherically symmetric BSs, and take the representative \texttt{A044} model. To assess numerical convergence, we compare the visibility amplitudes obtained with different set-ups of the adaptive mesh. The coarsest level is initialised at $100\times100$ pixels: if we scale the image size to be $100\times100$ $\mu$as (which can always be done by varying the scalar mass $\mu$), the resolution at the base level is $\Delta \theta = 1$ $\mu$as per pixel. With three (four, five) refinement levels, we thus get a resolution of $\frac{1}{8}$ $\left(\frac{1}{16}, \frac{1}{32}\right)$ $\mu$as: this corresponds to probing baselines up to $\lesssim 817$ (1634, 3268) G$\lambda$. We consider the \texttt{A044} model with three different setups: 1) the base setup with four refinement levels, and a maximum of $10^6$ geodesics integrated, 2) four refinement levels and as many geodesics as the AMR algorithm proposes ($\sim2.5 \times 10^6$ with default settings) and 3) five refinement levels and up to $4\times10^6$ geodesics. We plot the obtained VAs in Fig.~\ref{fig: resolution BS}, as well as the absolute differences $|(2)-(1)|$ and $|(3)-(1)|$. This suggests that, overall, variation is at the $\lesssim 10^{-3}$ level at large baselines, going down to about $\lesssim 5 \times 10^{-4}$ in the regime where we extract the modulations. This provides an estimate of the robustness of the obtained VAs.

\begin{figure}
    \centering
    \includegraphics[width=
    \linewidth]{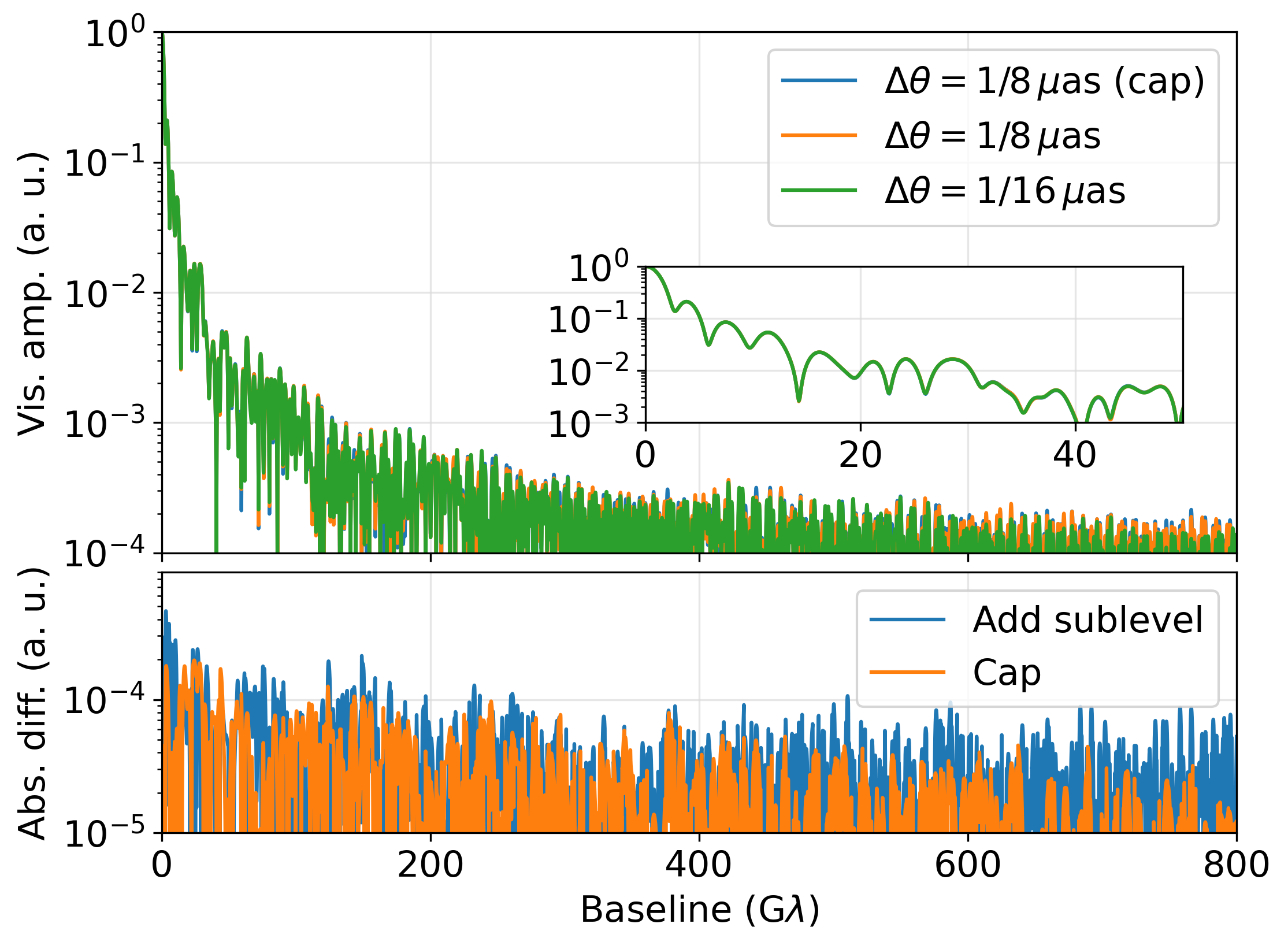}
    \caption{Comparison of the normalised visibility amplitudes obtained at different resolutions, for the \texttt{C38} rotating BS with the MSCO disc, at an inclination of 150$^\circ$. The image size is $100\times100$ $\mu$as, and $\Delta \theta$ is the angular separation between neighbouring pixels. (\textit{top}) We plot the visibility amplitude for 1) a low-resolution simulation, capped at $5\times10^5$ geodesics, 2) the same simulation without a cap on the number of geodesics and 3) a simulation with doubled resolution, capped at $2\times10^6$ geodesics. (\textit{bottom}) We plot the absolute differences between the simulations at different resolution, i.e.~adding a sublevel, and the difference between the low-resolution simulations with and without a cap.}
    \label{fig: resolution rot BS}
\end{figure}

We repeat the analysis for the \texttt{C38} model, to see if our results for rotating BS metrics can be trusted to the same extent as their non-rotating counterparts. 
We consider the \texttt{C38} model with three different setups: 1) three refinement levels, and a maximum of $5\times10^5$ geodesics integrated, 2) three refinement levels and as many geodesics as the AMR algorithm proposes ($\sim6.3 \times 10^5$ with default settings) and 3) four refinement levels and up to $2\times10^6$ geodesics. We plot the obtained VAs in Fig.~\ref{fig: resolution rot BS}, as well as the absolute differences $|(2)-(1)|$ and $|(3)-(1)|$. The results suggest that variation is at the $\lesssim 10^{-4}$ level.

\section{Additional rotating boson stars}

In this section, we provide some additional plots that compare the rotating BS models in Table \ref{tab: BS models} to their Kerr counterpart (i.e.~matching the mass and angular momentum).\\

\begin{figure*}
    \centering
    \includegraphics[width=0.7\linewidth]{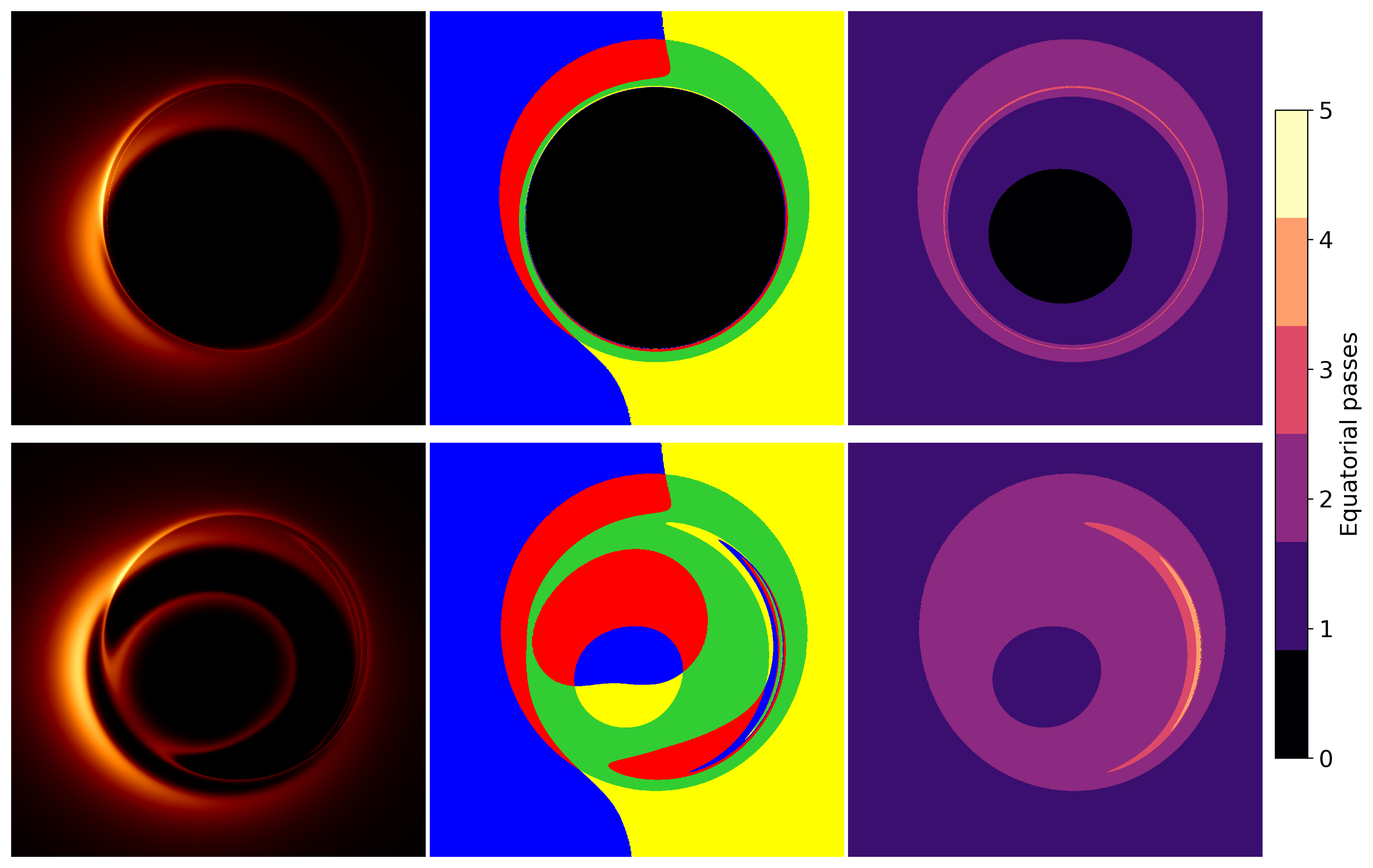}
    \caption{Comparison of a (\textit{top}) Kerr BH and (\textit{bottom}) the \texttt{C33} BS model at an inclination of 150\degree with respect to the rotation axis. From left to right, we show the observed intensity distribution, the four-color screen and the number of equatorial passes.}
    \label{fig: C33 comparison}
\end{figure*}

Figure \ref{fig: C33 comparison} compares the \texttt{C33} model, the least compact rotating BS we consider, with its Kerr counterpart. Immediately, one can see that (at this inclination) even the first photon ring no longer resembles that of Kerr: in the bottom left, it is no longer circular, and appears to connect with the inner lensed image. Visualizing the number of equatorial passes, we see that only a small set of geodesics experience sufficiently strong lensing to pass more than 2 times through the equatorial plane.

\begin{figure*}
    \centering
    \includegraphics[width=0.7\linewidth]{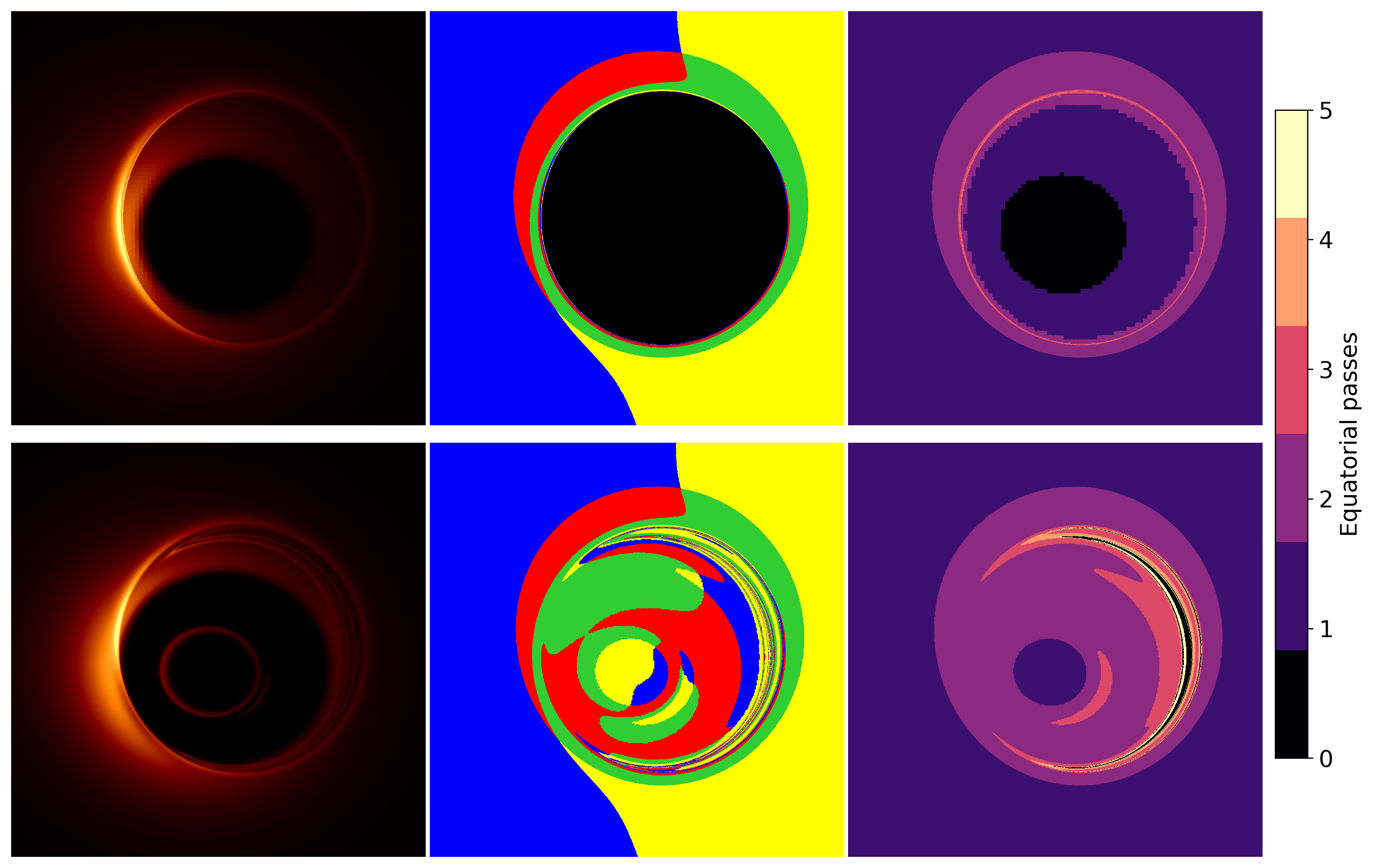}
    \caption{Same as Fig.~\ref{fig: C33 comparison}, but for the \texttt{C39} BS model. Note that, like in Fig.~\ref{fig: rot BS comparison}, the color scale for the number of equatorial passes has been capped at 5 for the BS, to facilitate comparison: the black arc in this image correspond to geodesics that have a higher number of equatorial passes, rather than an event horizon (as is the case for the BH).}
    \label{fig: C39 comparison}
\end{figure*}

\begin{figure*}
    \centering
    \includegraphics[width=0.7\linewidth]{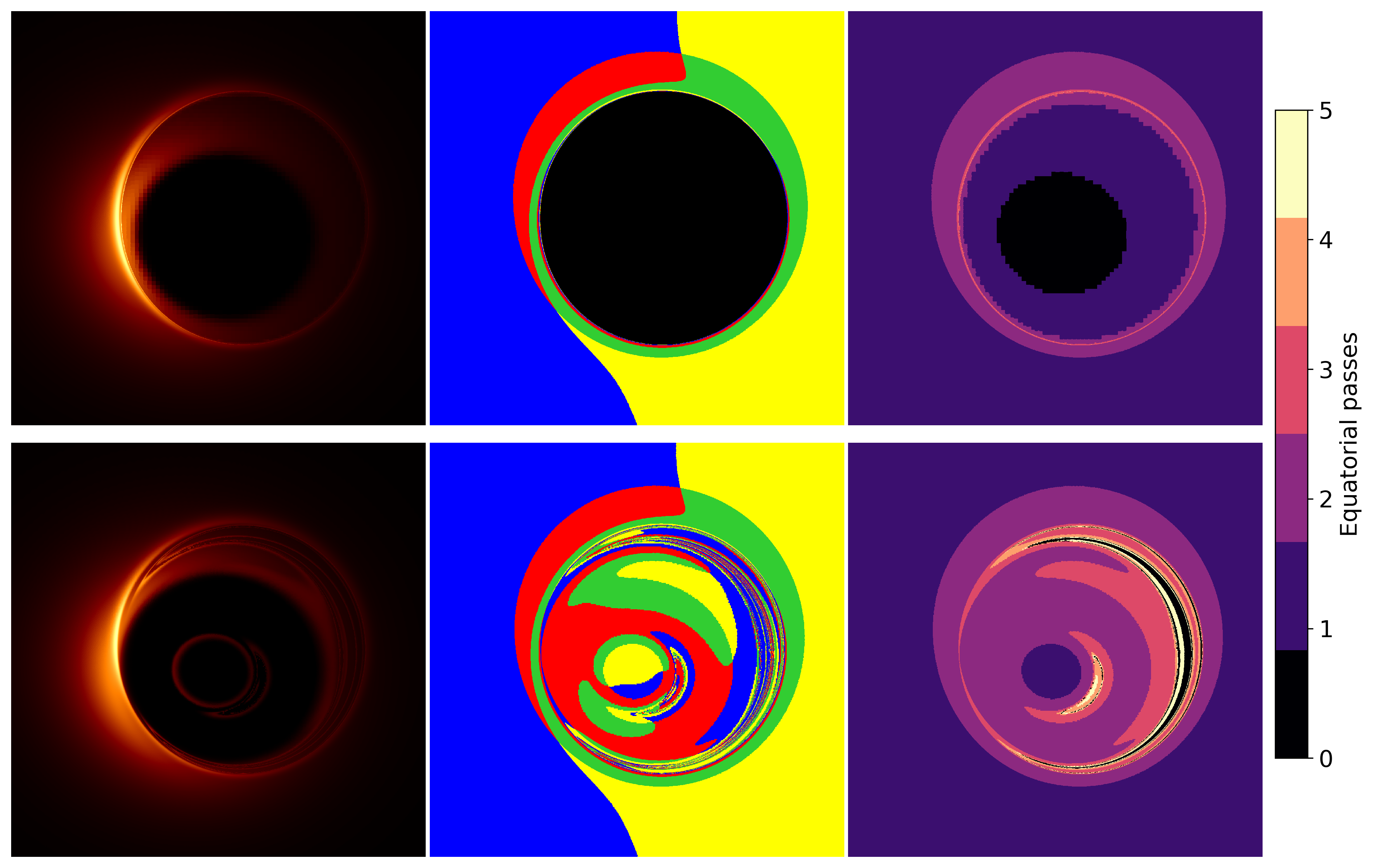}
    \caption{Same as Fig.~\ref{fig: C39 comparison}, but for the \texttt{C41} BS model.}
    \label{fig: C41 comparison}
\end{figure*}

Figures \ref{fig: C39 comparison} and \ref{fig: C41 comparison} show the analogue comparison for the \texttt{C39} and \texttt{C41} models: the interpretation follows that of the main text for Fig.~\ref{fig: rot BS comparison}, although these models are slightly more compact.

\section{Central disc model}
\label{app: central disc model}

As BSs do not have an event horizon, matter that gets accreted inwards does not simply disappear, but can be expected to accumulate in the core of the BS. To this end, e.g.~Refs.~\cite{Rosa:2023qcv, Li:2025awg} also considered a \textit{central disc model} for spherically symmetric BSs, described by \eqref{eq: GLM profile} with $\gamma = \rho = 0$ and $\sigma = 2M$. In this section, we briefly comment on the spherically symmetric case, and provide some images for the rotating BSs.\\

Ref.~\cite{Rosa:2023qcv} does not provide a (Keplerian) recipe for the fluid velocity, and hence their images are not prone to Doppler beaming. In principle, this is possible in \texttt{FOORT}, but implementing this correctly is left for future work, as a Keplerian disc that extends all the way to the core of an ultracompact spherical BS is not well described by the prescription in App.~\ref{app: FOORT}: the geodesic flow should be smoothly extended inwards from the MSCO according to the Cunningham prescription \cite{Cunningham:1975zz}, but revert to the circular flow inside the stable LR. --- see footnote \ref{foot: flow matching}. Once implemented, we expect the resulting images to be comparable to those in \cite{Rosa:2023qcv}, albeit with significant asymmetry from Doppler beaming for observers close to the equatorial plane.\\

For the rotating BSs, however, we can obtain a reasonable analogue to this central disc model. We can model a co-rotating central disc by extending the flow inwards from the prograde ISCO (see Fig.~\ref{fig: radial structure both}), and assuming Keplerian orbits for the fluid at all larger radii\footnote{The reason we cannot do this for the spherically symmetric BS is that no Keplerian orbits exist between the LRs.}. A more correct prescription would also extend the flow inwards from the MSCO, as long as the TCOs are unstable\footnote{One may also have to think about how to match this smoothly extended flow, which assumes the conserved quantities of the MSCO, to the marginally stable TCO between the LRs.}. This is also left for future work.

\begin{figure}
    \centering
    \includegraphics[width=\linewidth]{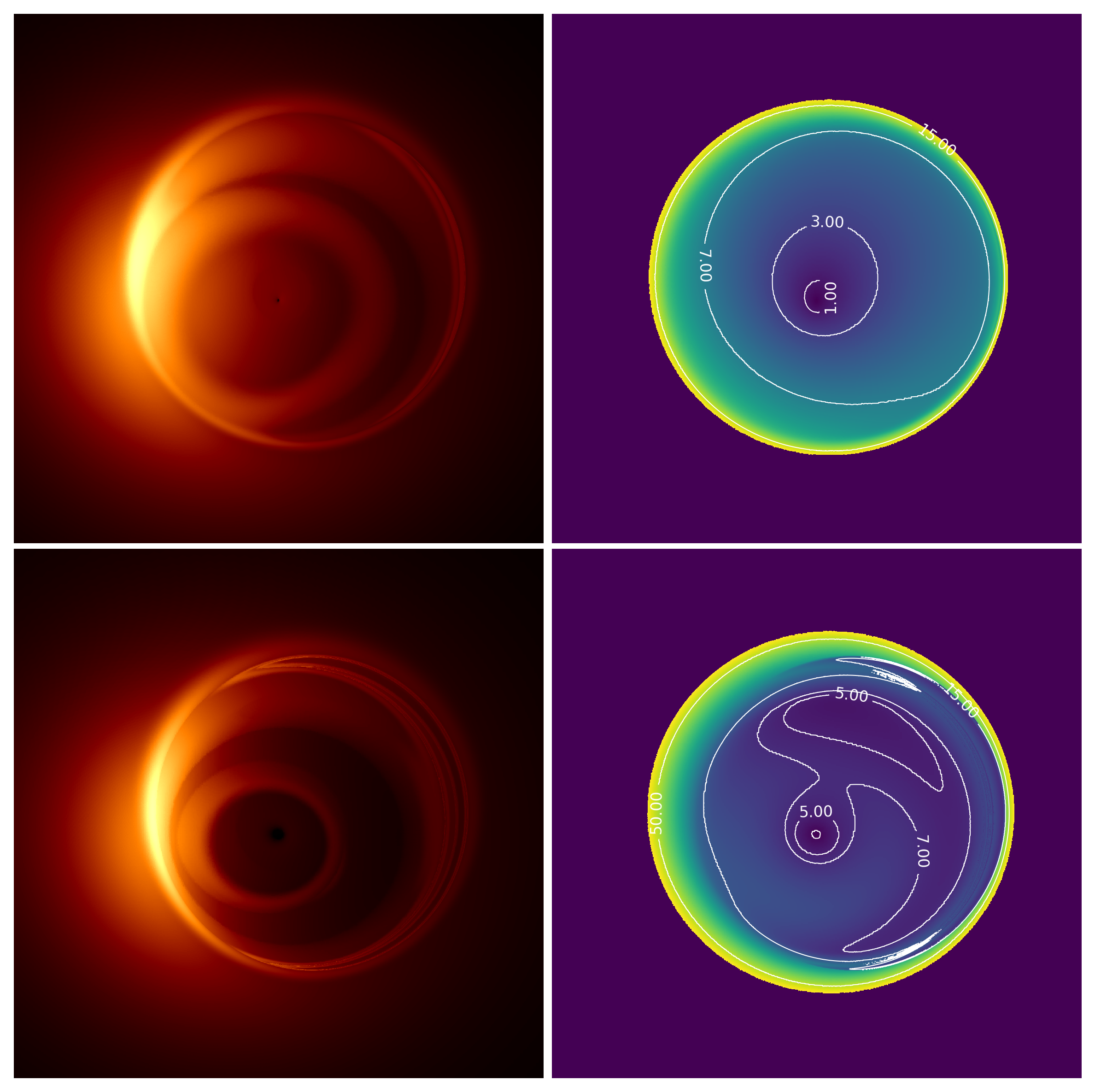}
    \caption{Ray-traced images of the (\textit{top}) \texttt{C33} and (\textit{bottom}) \texttt{C38} BSs with the central disc model of App.~\ref{app: central disc model}. The panels on the right show the minimal radius experienced \eqref{eq: r min} for geodesics in the central part of the image (regions where $\mu r_{\rm min}$ is larger than 16.5 and 56 respectively are truncated, to facilitate visualization), including several isocontours.}
    \label{fig: rot BS central disc}
\end{figure}

We provide images of the \texttt{C33} and \texttt{C38} models with the above central disc model in Fig.~\ref{fig: rot BS central disc}. Additionally, we visualize the smallest radius $r_{\rm min}$ that the geodesics experience in the right panels:
\begin{equation}
    \label{eq: r min}
    r_{\rm min} = \min_{\lambda \in \mathbb{R}} r(\lambda)\,,
\end{equation}
where $\lambda$ is an affine parameter. For \texttt{C33} the central brightness depression is nearly completely gone, whereas \texttt{C38} still boasts a relatively dark central region (aside from the ring-shaped structure that is always present). We see that the isocontours of $r_{\rm min}$ for \texttt{C33} are somewhat concentric circles, whereas \texttt{C38} has two regions on the image plane in which geodesics get close to the BS center. For the latter, we also see the crescent-shaped structures appear in the isocontour corresponding to $\mu r = 15$.

\bibliography{ref}

\end{document}